# Diffraction-Unlimited Tip-Enhanced Sum-Frequency Vibrational Nanoscopy

*Shota Takahashi[1], Koichi Kumagai[2], Atsunori Sakurai[1,3]\*, Tatsuto Mochizuki[1,3], Tomonori Hirano[2], Akihiro Morita[2]\*, and Toshiki Sugimoto[1,3]\**

[1]Department of Materials Molecular Science, Institute for Molecular Science (IMS), Okazaki, Aichi 444-8585, Japan

[2]Department of Chemistry, Graduate School of Science, Tohoku University, Sendai, Miyagi 980-8578, Japan.

[3]The Graduate University for Advanced Studies, SOKENDAI, Okazaki, Aichi, 444-8585, Japan

\*Corresponding authors
E-mail: asakurai@ims.ac.jp, morita@tohoku.ac.jp, toshiki-sugimoto@ims.ac.jp,

ABSTRACT: Sum-frequency generation (SFG) is a powerful second-order nonlinear spectroscopic technique that provides detailed insights into molecular structures and absolute orientations at surfaces and interfaces. However, conventional SFG based on far-field schemes suffers from the diffraction limit of light, which inherently averages spectroscopic information over micrometer-scale regions and obscures nanoscale structural inhomogeneity. Here, we overcome this fundamental limitation by leveraging a highly confined optical near field within a tip–substrate nanogap of a scanning tunneling microscope (STM), pushing the spatial resolution of SFG down to ~10 nm, a nearly two-orders-of-magnitude improvement over conventional far-field SFG. By capturing tip-enhanced SFG (TE-SFG) spectra concurrently with STM scanning, we demonstrate the capability to resolve nanoscale variation in molecular adsorption structures across distinct interfacial domains. To rigorously interpret the observed TE-SFG spectra, we newly developed a comprehensive theoretical framework for the TE-SFG process and confirm via numerical simulations that the TE-SFG response under our current experimental conditions is dominantly governed by the dipole-field interactions, with negligible contributions from higher-order multipole effects. The dominance of the dipole mechanism ensures that the observed TE-SFG spectra faithfully reflect not only nanoscale interfacial structural features but also absolute up/down molecular orientations. This study presents the first experimental realization of diffraction-unlimited second-order nonlinear vibrational SFG nanoscopy, opening a new avenue for nanoscale domain-specific investigation of molecular structures and dynamics within inhomogeneous interfacial molecular systems beyond the conventional far-field SFG and STM imaging.

**TOC graphic**

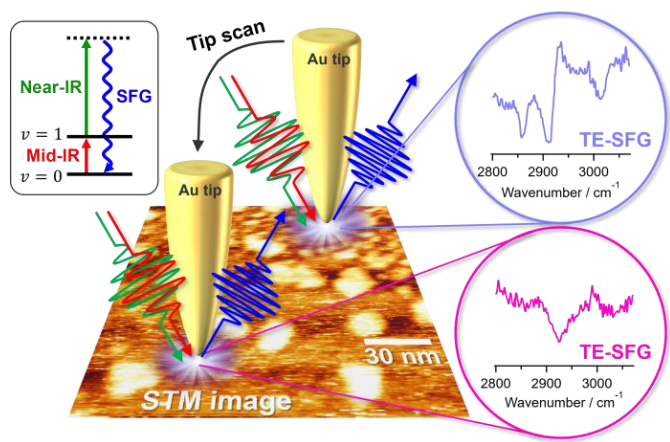

**Introduction**

Vibrational sum-frequency generation (SFG, Figure 1a) is a second-order nonlinear spectroscopic technique that has proven to be powerful for characterizing molecular structures and dynamics at surfaces and interfaces.[1–4] Owing to the symmetry constraints, the SFG process is forbidden in centrosymmetric bulk materials, restricting its optical activity to the interfacial regions between adjacent media. A particularly notable characteristics of SFG is its ability to distinguish the absolute up/down orientation of interfacial molecules,[3,5–9] a capability inaccessible via linear vibrational spectroscopy such as infrared absorption and Raman scattering. The inherent interface specificity and molecular orientation sensitivity of SFG are key attributes that have led to its widespread utilization in investigating diverse interfacial phenomena, such as photochemical reactions at solid and liquid surfaces,[10,11] charge transfer at organic and inorganic interfaces,[12] and ferroelectric crystal growth on metallic surfaces.[6,13,14] However, a major limitation of conventional far-field SFG lies in its diffraction-limited spatial resolution. While efforts to observe microstructures using SFG combined with optical microscope are actively ongoing,[15–19] the maximum achievable resolution is restricted to the micrometer-scale due to the fundamental diffraction limit of light. Consequently, the spectroscopic information obtained through vibrational SFG is statistically averaged over a vast number ($>10^6$) of chemical species or molecules, restricting its applicability to highly inhomogeneous molecular systems where nanoscale spatially resolved spectroscopic insights are essential.

One promising strategy to overcome the limitation of spatial resolution is to exploit plasmonic near-field confinement and enhancement of light within the nanogap formed between a metallic tip and substrate in scanning probe microscopes, such as an atomic force microscope (AFM) and a scanning tunneling microscope (STM).[20] Such confined near-field light amplifies various optical processes in the nanometric region directly beneath the AFM/STM tip, providing unique platforms for nanoscale spectroscopic imaging of both linear[21–23] and nonlinear[24–26] optical properties. Early demonstrations of near-field nonlinear optical mapping employed metal-coated optical fibers as aperture-type probes to collect nonlinear optical signals with a spatial resolution of ~100 nm.[27,28] However, to achieve higher spatial resolution on the order of 10 nm, it is essential to harness the intense field confinement and enhancement via gap-mode plasmons, which are excited at the nanogap between an apertureless metallic tip apex and substrate.[29–31] Indeed, by leveraging the plasmonic tip-enhancement effect to amplify nonlinear optical signals, including second-harmonic generation (SHG), SFG, two-photon photoluminescence, and four-wave mixing (FWM), nonlinear nanoimaging has been successfully demonstrated for plasmonic metal nanostructures,[24,25,32–34] ferroelectric materials,[35,36] quantum dots,[37] and transition-metal dichalcogenides.[38–40]

Since these nonlinear optical processes involve frequency conversion between input and output light, efficient tip-enhancement requires the simultaneous amplification of the energetically separated input and output light. Thus, for tip-enhanced nanoimaging based on vibrational SFG, both molecular vibrationally resonant mid-infrared (IR) light ($\omega_{MIR}$ in Figure 1a) and higher-energy near-IR/visible light ($\omega_{NIR}$ and $\omega_{SFG}$ in Figure 1a) must be enhanced concurrently. However, the gap-mode plasmon resonances typically occur in the visible to near-IR region,[20,41–45] preventing efficient enhancement of vibrational SFG process. Although micrometer-scale plasmonic structures can enhance fields in the mid-IR regions,[46–48] their relatively large structures typically weaken the enhancement of near-IR/visible light. Consequently, achieving broadband field enhancement encompassing visible-to-mid-IR range remains highly challenging.

In this paper, we present the first demonstration of tip-enhanced SFG (TE-SFG) vibrational nanoscopy with spatial resolution down to ~10 nm, representing a 100-fold improvement over micrometer-scale diffraction-limited spatial resolution of conventional far-field SFG. A key strategy to overcome the limited spectral range of the field enhancement is the simultaneous utilization of gap-mode plasmon resonances within the tip–substrate nanocavity and the antenna effect of the micrometer-scale tip shaft, a combination that we recently demonstrated to be effective.[49,50] The antenna effect enables strong field enhancement of the vibrationally resonant mid-IR light ($\omega_{MIR}$) and the subsequent near-IR light ($\omega_{NIR}$), whereas the upconverted SFG light ($\omega_{SFG}$), induced by these two-color excitation pulses, can be enhanced through the gap-mode plasmon resonance. The simultaneous enhancement of these input and output light enables the enhancement of the overall SFG signals from the nanogap.[49,50] In this study, we exploit this combined enhancement mechanism to capture vibrational TE-SFG spectra that resolve domain-level structural variations with ~10 nm spatial resolution in a spatially inhomogeneous molecular monolayer formed on a Au(111) substrate. Furthermore, by analyzing the optical interference between vibrationally non-resonant and resonant TE-SFG signals, we demonstrate that the molecular vibrational nonlinear susceptibility ($\chi^{(2)}$) spectra of spatially distinct nanometric domains can be directly extracted. To rigorously interpret the obtained TE-SFG spectra and establish this technique as a reliable probe for interfacial molecular structures, we also developed a comprehensive theoretical framework that incorporates dipole–field and quadrupole–field-gradient interactions specific to the near-field regime. Unlike far-field SFG,[51–55] where the electric fields are treated as plane waves, near-field excitation involves strongly confined and spatially varying fields. To take account of these characteristics, we carried out numerical electromagnetic simulations of the field distributions within the nanogapand combined them with quantum chemical calculations. Our analysis revealed that the dipole approximation remains valid under our experimental conditions, supporting that the extracted $\chi^{(2)}$ spectra reliably reflects the absolute up/down molecular orientation. This indicates that a unique characteristic of second-order nonlinear optical responses, namely their intrinsic sensitively to interfacial polarity, is preserved even in the near-field scheme, validating the potential of TE-SFG as a powerful probe for structural features and absolute up/down molecular orientation at the nanoscale.

**Results and Discussion**

**TE-SFG measurements**

As a platform for the demonstration of diffraction-unlimited TE-SFG measurements, we employed a near-field nonlinear spectroscopic system combining an STM unit with a mid- and near-IR wavelength-tunable pulse laser system (Figure 1b).[50] A smooth and sharpened home-made plasmonic Au tip[56] with the apex of ~50-nm radius of curvature (Figure 1c) was mounted in the STM unit. A self-assembled monolayer (SAM) of 4-methylbenzenthiol (4-MBT, Figure 1d) on an atomically flat Au(111) substrate was prepared as a model sample for the TE-SFG measurements. Since 4-MBT molecules adsorb via a covalent bond between the sulfur atom and the gold atom, the methyl group located at the opposite end of the molecule naturally points upward relative to the surface (Figure 1b).[57,58] As discussed later, the well-defined molecular orientation of the 4-MBT SAM model system allows us to assess whether the TE-SFG signals accurately reflects the information on the absolute up/down orientation of the molecules. A topographic image of 4-MBT SAM on Au(111) surface is shown in Figure 1e. A number of protrusion structures that can be seen in the image are Au adatom islands, which is a typical characteristic of SAMs of arenethiols on Au(111).[50,59,60] These island structures arise from the well-known phenomena of surface reconstruction of Au atoms induced by the adsorption of arenethiol molecules,[50,59,60] and thus molecules are adsorbed both on the islands and surrounding regions.[59] Therefore, the observation of such island structures indicates the formation of a single monolayer of 4-MBT molecules.

We begin by examining the spectra obtained from a bare Au substrate as a reference spectrum. To induce the TE-SFG process, we used spatially and temporally overlapped *p*-polarized laser pulses: a wavelength-tunable mid-IR pulse (scan range 2800–3050 cm$^{-1}$, 300 fs, FWHM: 70 cm$^{-1}$, 400 pJ) and a narrow-band near-IR pulse (9674.0 cm$^{-1}$ (1033.7 nm), 1 ps, FWHM: 10 cm$^{-1}$, 10 pJ). The spectra of these excitation pulses are provided in Figures S1a and b. These weak laser pulses were focused onto the tip-substrate nanogap at a repetition rate of 50 MHz, and the forward-scattered TE-SFG signals were collected in reflection geometry (Figure 1b). The far-field SFG signal observed when the substrate was retracted from the tip by ~30 nm (solid black curve in Figure 1f) was negligibly small under these weak irradiation conditions. Although tip plasmons localized at the tip apex can still be excited even at this separation, their contribution to the SFG signal is negligible, as confirmed by our numerical electromagnetic field simulation (Figure S6). In contrast, when the tip approached the substrate under a sample bias of 0.1 V, the SFG signal intensity dramatically increased (broken red curve in Figure 1f). We confirmed that contributions from DC-field-induced third-order SFG[61–63] across the nanogap are negligible under our small bias application of 0.1 V. Therefore, the enhanced signal is attributed to the second-order TE-SFG, arising from the vibrationally non-resonant $\chi^{(2)}$ optical response of surface electrons of the Au tip and substrate:[49,50]

$$I_{\text{TESFG}} \propto \left|\chi^{(2)}_{\text{NR}}\right|^2 I_{\text{NIR}} I_{\text{MIR}}, \quad (1)$$

where $I_{\text{TESFG}}$ is the output TE-SFG signal intensity; $\chi^{(2)}_{\text{NR}}$ is the vibrationally non-resonant second-order nonlinear susceptibility of the Au tip and substrate; and $I_{\text{NIR}}$ and $I_{\text{MIR}}$ are the intensities of the near-IR and mid-IR fields enhanced within the nanogap, respectively. The near-IR field intensity within the gap $I_{\text{NIR}}$ is given by the product of incident field intensity ($I_{\text{NIR0}}$) and the field enhancement factor within the gap ($K_{\text{gap}}$): $I_{\text{NIR}} = \left|K_{\text{gap}}(\omega_{\text{NIR}})\right|^2 I_{\text{NIR0}}$.[49] The mid-IR field intensity within the gap $I_{\text{MIR}}$ can also de described in a similar way: $I_{\text{MIR}} = \left|K_{\text{gap}}(\omega_{\text{MIR}})\right|^2 I_{\text{MIR0}}$.[49] As shown in Figure S5b, the field enhancement factor $K_{\text{gap}}$ can be regarded as almost a constant value in the scan range of mid-IR wavenumber (2800–3050 cm$^{-1}$). Therefore, the observed Gaussian-like feature-less spectral shape (red broken curve in f) reflects the frequency distribution of the convolution of mid-IR and near-IR excitation pulses (Figure S1c).

In contrast, the TE-SFG spectrum of the 4-MBT-adsorbed Au substrate exhibited a distorted spectral feature (solid red curve in Figure 1f). This distortion is attributed to the interference between the vibrationally non-resonant TE-SFG signal from Au and the vibrationally resonant TE-SFG signal from molecules in the nanogap:

$$I_{\text{TESFG}} \propto \left|\chi_{\text{NR}}^{(2)} + \chi_{\text{R}}^{(2)}\right|^2 I_{\text{NIR}} I_{\text{MIR}} = \left|\chi_{\text{total}}^{(2)}\right|^2 I_{\text{NIR}} I_{\text{MIR}}, \quad (2)$$

where $\chi_{\text{R}}^{(2)}$ is the vibrationally resonant second-order nonlinear susceptibility of molecules in the nanogap and $\chi_{\text{total}}^{(2)} = \chi_{\text{NR}}^{(2)} + \chi_{\text{R}}^{(2)}$. The frequency distribution of the excitation pulses ($I_{\text{NIR}}$ and $I_{\text{MIR}}$ in eqs. (1) and (2)) can be eliminated by taking the ratio of the intensity spectra with 4-MBT molecules (eq. (2), solid red curve in Figure 1f) and without 4-MBT molecules (eq. (1), broken red curve in Figure 1f) in the nanogap, yielding the $\left|\chi_{\text{total}}^{(2)}\right|^2$ spectrum of the molecule-embedded nanogap normalized by $\left|\chi_{\text{NR}}^{(2)}\right|^2$ (Figure 1g):

$$\left|\chi_{\text{total}}^{(2)}\right|_{\text{norm}}^2 \equiv \left|\frac{\chi_{\text{total}}^{(2)}}{\chi_{\text{NR}}^{(2)}}\right|^2 = \left|1 + e^{-i\phi_{\text{NR}}} \frac{\chi_{\text{R}}^{(2)}}{\left|\chi_{\text{NR}}^{(2)}\right|}\right|^2, \quad (3)$$

where $\left|\chi_{\text{NR}}^{(2)}\right|$ and $\phi_{\text{NR}}$ represent the absolute value and phase of the vibrationally non-resonant susceptibility $\chi_{\text{NR}}^{(2)}$, respectively. As discussed later, the analysis of $\left|\chi_{\text{total}}^{(2)}\right|_{\text{norm}}^2$ allows for the characterization of molecular vibrational modes in the nanogap.

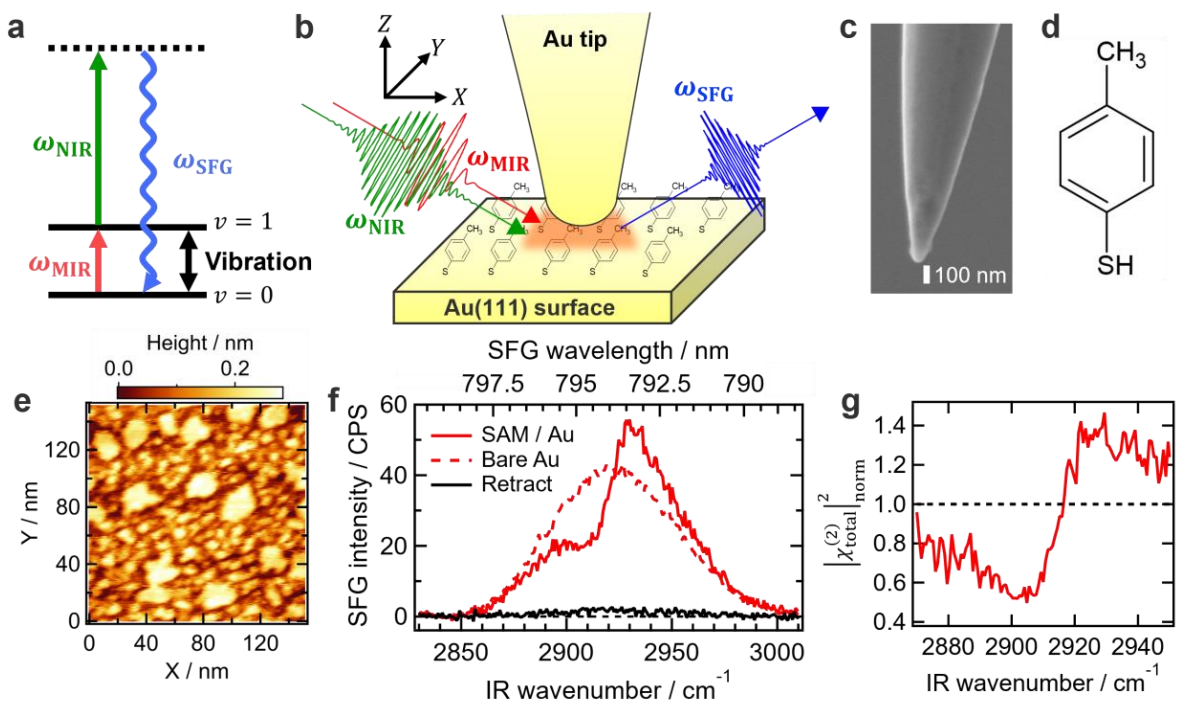

Figure 1. (a) Energy diagram of vibrationally resonant SFG. (b) Schematic depiction of TE-SFG experiment using a Au tip and SAM-covered Au substrate, where *p*-polarized mid- and near-IR pulses were used as excitation sources. The radiated signal was detected in the reflection geometry. (c) Scanning electron micrograph of the Au tip used in the experiments. (d) Molecular structure of 4-MBT. (e) STM image of 4-MBT SAM measured at a sample bias of 0.1 V and a setpoint current of 100 pA. The presence of a number of protrusions is a typical characteristic of arenethiol SAMs attributed to Au adatom islands rather than to bumps of molecular aggregates, with the 4-MBT monolayer formed not only on the wide terrace region but also on the islands.[50,59,60] (f) TE-SFG spectra obtained for a bare Au substrate (broken red curve) and a 4-MBT SAM/Au substrate (solid red curve). The sample bias and tunneling current setpoint were 0.1 V and 100 pA, respectively. The black curve indicates the signal obtained when the tip and the substrate were retracted by 30 nm, sufficiently reducing plasmonic enhancement effects. The pulse energies of the mid-IR and near-IR excitation light were 400 pJ and 10 pJ, respectively. (g) Spectrum of $\left|\chi^{(2)}_{total}\right|^2_{norm}$ obtained using the TE-SFG results shown in (f) and eq. (3) in the main text.

**Domain-resolved TE-SFG spectra and comparison with FF-SFG**

As already shown in Figure 1e, 4-MBT SAM on the Au(111) surface typically possesses a number of island structures (Au adatom islands). While this island-rich structure occupies a predominantly large area on the surface, careful STM scanning over a wide region revealed the presence of localized domains with a lower island density (Figure S2). Such a difference in the nanoscale surface topography suggests different adsorption structures of 4-MBT molecules, which may cause differences in TE-SFG spectra across these domains. Thus, to evaluate a spatial resolution of TE-SFG nanoscopy, we focused on these two types of nanoscale domains and conducted TE-SFG measurements across the boundary region between those two domains (Figure 2a). Since the island-rich region is dominant on the surface (Figure S2a), we hereafter refer to this island-rich region (upper half of Figure 2a) as the "major domain" and the island-less region (lower half of Figure 2a) as the "minor domain."

TE-SFG measurements were carried out while simultaneously performing sample scanning, which enabled us to avoid the optical damage accumulation caused by fixing the tip at the same position. Specifically, we performed a STM scan over the 155 nm × 155 nm region for 100 seconds under the irradiation of excitation light for TE-SFG, during which TE-SFG spectra were acquired in ten consecutive measurements. The accumulation time for each measurement was ten seconds. Then, we acquired ten TE-SFG spectra corresponding to ten partitioned regions (#1–10) in Figure 2a. As shown in Figure 2b, different spectral features were observed for the minor (#1–5) and major (#6–10) domains: the minor domain exhibited a dip structure centered at around 2925 cm$^{-1}$, whereas the major domain showed a dispersive spectral profile almost identical to that in Figure 1f. This indicates that the observed TE-SFG spectra captured the local structural differences in the molecular system beneath the nanoscale tip apex. We note that angstrom-scale Au monoatomic step structures were clearly observed in the STM image under excitation light irradiation (Figure S2). This ensures that optical damage to the tip and sample is negligible in our TE-SFG measurements.

The domain dependent change in the spectral shapes was well quantified in Figure 2c by plotting the variation of the TE-SFG intensity at around 2925 cm$^{-1}$ (red shaded region in Figure 2b). The intensity profile exhibited a transition at the boundary between minor and major domains (Figure 2c). The rise behavior of this profile reveals that a diffraction-unlimited spatial resolution of ~30 nm was achieved at the present stage. This represents stark contrast to the conventional far-field SFG measurements, whose spatial resolution is fundamentally constrained by diffraction-limited focus spot sizes on the order of micrometer scale.

Then, the TE-SFG intensity spectra for the domains #1–10 (Figure 2b) were converted into the $\left|\chi^{(2)}_{\text{total}}\right|^2_{\text{norm}}$ spectra (Figure 2d). As shown in Figure 2d, the minor (#1–5) and major (#6–10) domains exhibited different spectral shapes: the dip structure for the minor domain and the dispersive structure for the major domain. Such variation in the spectral shapes suggests difference in molecular adsorption structures in the minor and major domains. To verify the structural difference in more detail, we measured the $\left|\chi^{(2)}_{\text{total}}\right|^2_{\text{norm}}$ spectra of both domains over broader wavenumber range by systematically tuning the center wavenumber of mid-IR excitation pulses (Figure 3a and b, see Figure S1b for a series of the spectra of mid-IR excitation pulses). $\left|\chi^{(2)}_{\text{total}}\right|^2_{\text{norm}}$ spectra over broader wavenumber range (Figure 3a and b) exhibit multipeak structures, suggesting the observation of multiple vibrational modes in our TE-SFG measurements. When multiple vibrational modes contribute to the overall vibrationally resonant $\chi^{(2)}_{R}$, $\left|\chi^{(2)}_{\text{total}}\right|^2_{\text{norm}}$ is represented as:[2,7,57,64–70]

$$\left|\chi^{(2)}_{\text{total}}\right|^2_{\text{norm}} = \left|1 + e^{-i\phi_{\text{NR}}} \sum_a \frac{\chi^{(2)}_{R,a}}{\left|\chi^{(2)}_{\text{NR}}\right|}\right|^2 = \left|1 + e^{-i\phi_{\text{NR}}} \sum_a \frac{A'_a}{\omega_a - \omega_{\text{MIR}} - i\Gamma_a}\right|^2, \quad (4)$$

where $\chi^{(2)}_{R,a}$ is the second-order nonlinear optical susceptibility derived from the vibrational mode $a$, and the overall vibrationally resonant susceptibility $\chi^{(2)}_{R}$ is given by the summation of $\chi^{(2)}_{R,a}$ $\left(\chi^{(2)}_{R} = \sum_a \chi^{(2)}_{R,a}\right)$. $\omega_a$ and $\Gamma_a$ are the resonant frequency and damping constant of the vibrational mode $a$, respectively. $A'_a$ is the resonant amplitude of the vibrational mode $a$ normalized by the absolute value of vibrationally non-resonant susceptibility $\left(A'_a = A_a / \left|\chi^{(2)}_{\text{NR}}\right|\right)$. Fitting the $\left|\chi^{(2)}_{\text{total}}\right|^2_{\text{norm}}$ spectra in Figure 3a and b using eq. (4) yields the $\chi^{(2)}_{R,a}$ spectrum for each mode in both the major and minor domains, enabling detailed comparison of their structures. Free parameters in the fitting are $\phi_{\text{NR}}$, $A'_a$, $\omega_a$, and $\Gamma_a$. Previous studies have shown that $\phi_{\text{NR}}$ is in the range of $0 \leq \phi_{\text{NR}} \leq \frac{\pi}{2}$.[50,64,66,71] By resolving the $\pi$ arbitrariness of $\phi_{\text{NR}}$[72], both $\phi_{\text{NR}}$ and $A'_a$ values are uniquely determined. The resulting fits, shown as green curves in Figure 3a and b, closely match the experimental spectra, with the best-fit parameters summarized in Table S1. Based on the extracted parameters for the vibrationally resonant responses ($A'_a$, $\omega_a$, and $\Gamma_a$), we reconstructed the $\text{Im}(\chi^{(2)}_{\text{R},a})$ spectra, which directly represent the pure absorptive molecular resonances (Figure 3c and d).

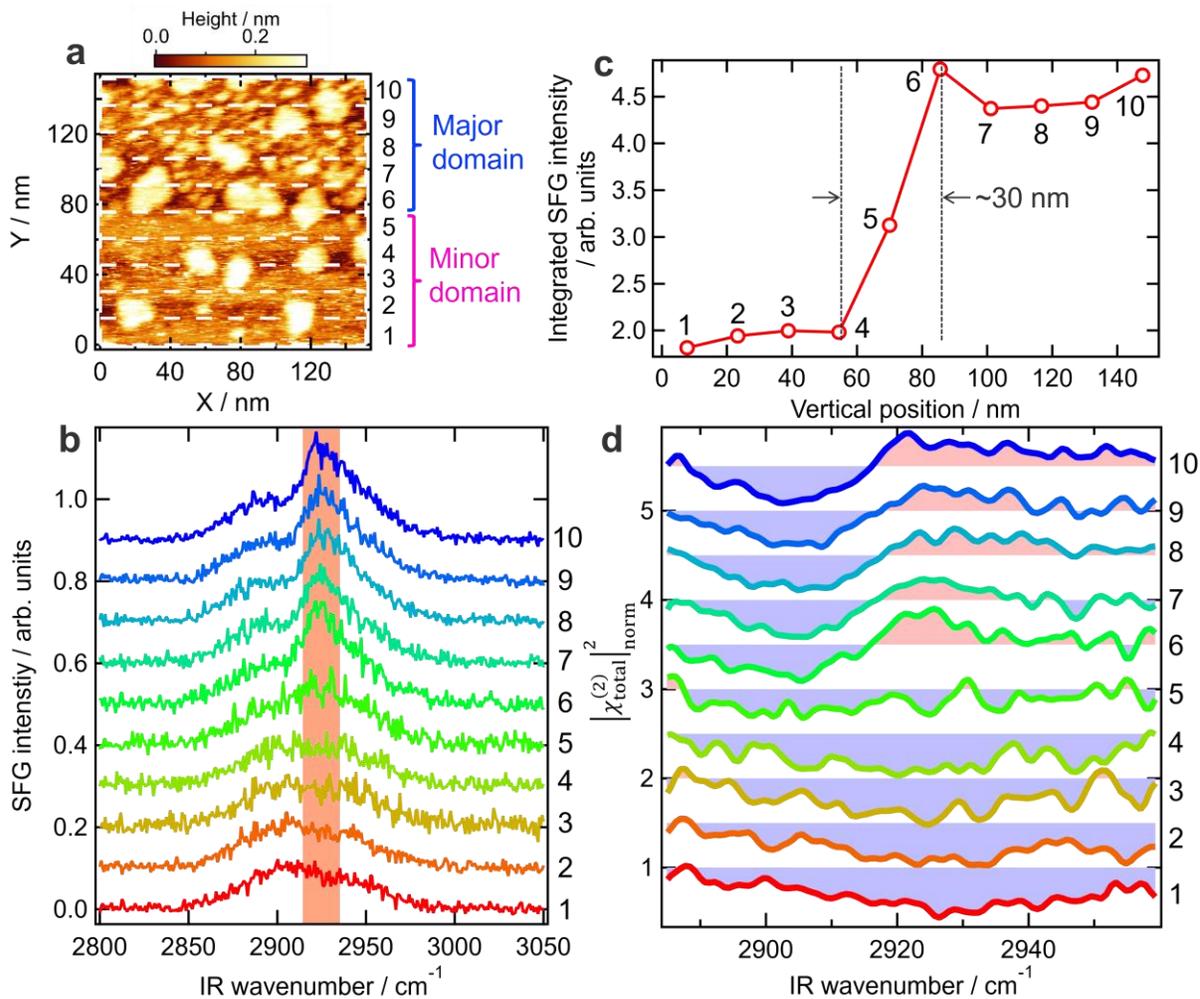

Figure 2. (a) STM image of 4-MBT SAM measured at a sample bias of 0.1 V and a setpoint current of 100 pA. The area with many protrusions in the upper half and that with few bumps in the lower half are denoted as "major domain" and "minor domain", respectively. The image was obtained while irradiating the STM nanogap by the excitation lasers for TE-SFG experiments. (b) Spectrally resolved TE-SFG signals emitted from the ten rectangular domains (#1–10) depicted in (a). (c) The plot of the integrated SFG intensities between 2917 and 2932 cm$^{-1}$, which are marked by the red shaded area in (b). (d) $\left|\chi^{(2)}_{\text{total}}\right|^2_{\text{norm}}$ spectra for ten rectangles in (a). Areas with a value greater than (less than) 1 are filled with red (blue) shade.

As shown in Figure 3c and d, four vibrational peaks (1)–(4) appear at nearly identical wavenumbers in both domains. Based on our quantum chemical calculations, these peaks are assigned as follows: (1, 2) Fermi resonance levels formed between the $CH_3$ symmetric stretching and the overtone of the $CH_3$ bending modes ($r^+_{\text{FR1}}$, $r^+_{\text{FR2}}$), (3) a pair of degenerate $CH_3$ asymmetric stretching modes ($r^-_1$, $r^-_2$), and (4) two degenerate C–H stretching modes in the phenyl ring ($r_{\text{ph1}}$, $r_{\text{ph2}}$). Further details of these assignments are provided in Supporting Information section 5.

SFG measurements were also performed in a far-field geometry (Figure S3), where the tip–sample distance was maintained at more than 1 µm to suppress the gap-mode plasmon excitation. In this configuration, the intensity of near-IR excitation light was increased by one order of magnitude compared to that used in TE-SFG measurements, allowing the acquisition of a detectable far-field SFG (FF-SFG) signal. A fitting analysis based on eq. (4) was also conducted for the FF-SFG spectrum (Figure S3a), yielding the parameters for the vibrationally resonant responses ($A'_a$, $\omega_a$, and $\Gamma_a$, see Table S1) and the corresponding $\text{Im}\left(\chi^{(2)}_{\text{R},a}\right)$ spectra for four vibrational modes (Figure S3b). The $\text{Im}\left(\chi^{(2)}_{\text{R}}\right)\left(=\sum_a \text{Im}\left(\chi^{(2)}_{\text{R},a}\right)\right)$ spectra are shown as broken curves in Figures 3c, 3d, and S3b. Notably, the TE-SFG spectrum of the major domain (Figure 3c) resembles the FF-SFG spectrum (Figure S3b), whereas the spectrum of the minor domain (Figure 3d) exhibits different spectral features. This difference arises from the spatial averaging nature of FF-SFG. Since FF-SFG collects signals from a micrometer-scale focal area encompassing both major and minor domains, the contribution from the sparsely distributed minor domains is overwhelmed by that of the major domain (Figure S2a). As a result, conventional FF-SFG measurements cannot capture the local spectral characteristics of the minor domain. In contrast, TE-SFG overcomes this limitation by confining the excitation field to a nanometric region, enabling the selective detection of local optical responses. The successful extraction of spectral features from the minor domain, which are otherwise hidden in the ensemble-averaged FF-SFG signal, highlights the advantage of TE-SFG for nanoscopic investigation of heterogeneous surface molecular systems.

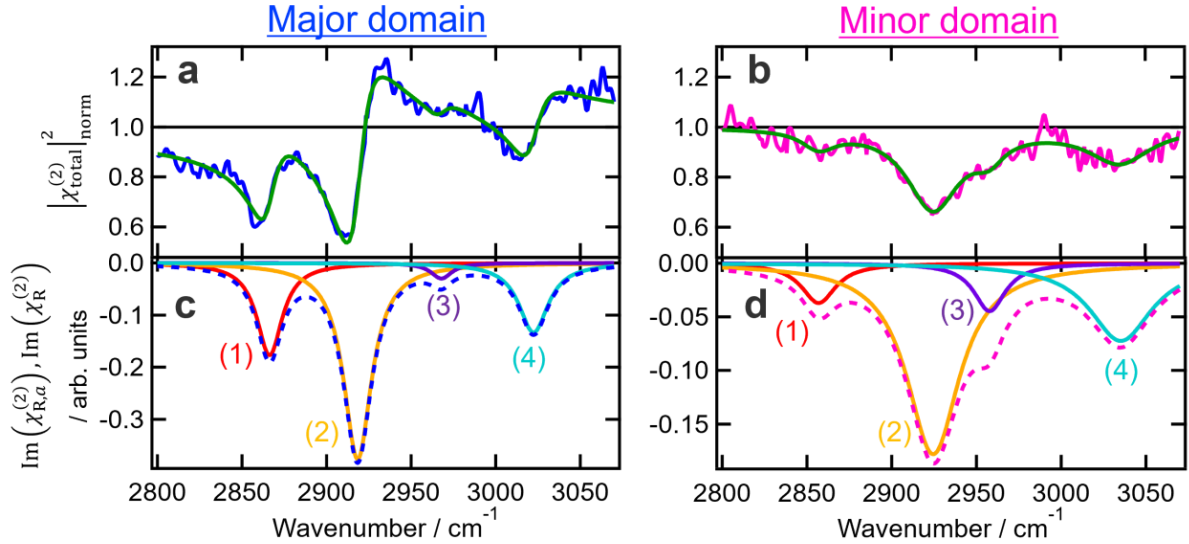

Figure 3. (a, b) $|\chi_{\text{total}}^{(2)}|^2_{\text{norm}}$ spectra in the CH stretching region of 4-MBT molecules. Green curves are the results of curve fitting based on eq. (4). (c, d) Spectra of $\text{Im}(\chi_{R,a}^{(2)})$ corresponding to four different vibrational modes of a 4-MBT molecule. The left (a and c) and right (b and d) columns represent the results of the TE-SFG measurements for the major domain and the minor domain, respectively. Broken curves in c and d are the $\text{Im}(\chi_R^{(2)})$ spectra obtained by summing the four $\text{Im}(\chi_{R,a}^{(2)})$ spectra.

**Theoretical framework for TE-SFG mechanisms: dipolar and quadrupolar contributions**

To gain physical insights into $\text{Im}(\chi_R^{(2)})$ spectra obtained from the TE-SFG measurements (Figures 3c and d), we developed a theoretical framework describing microscopic mechanisms underlying the TE-SFG process. In contrast to the tip-enhanced Raman scattering (TERS) or IR scattering-type scanning near-field optical microscopy (IR-*s*-SNOM), TE-SFG is a relatively new approach[50] and a well-established theoretical description has yet to be developed. To address this, we formulated a theoretical model that accounts for possible SFG mechanisms including dipole–field interaction and higher-order multipolar contributions. Then, we quantitatively evaluate relative contributions of these interactions under the current experimental condition through electromagnetic field simulations and quantum chemical calculations.

SFG from molecules is regarded as radiation from molecular nonlinear optical polarizations induced by two successive interactions between the electric field and the molecular system (Figure 1a). When the spatial variation of the electric field is negligible compared to molecular dimensions, the light–molecule interaction is dominated by the coupling between the molecular dipole moment and the electric field ($-\hat{\boldsymbol{\mu}} \cdot \boldsymbol{E} = -\sum_i^{x,y,z} \hat{\mu}_i E_i$). However, under conditions where the electric field gradient becomes comparable to molecular dimensions,[73,74] higher-order multipolar contributions may become non-negligible and should be considered to more accurately describe the light-molecule interaction beyond the dipole approximation:

$$\hat{V} = -\hat{\boldsymbol{\mu}} \cdot \boldsymbol{E} - \hat{\boldsymbol{q}}{:}\nabla \boldsymbol{E}, \tag{5}$$

where $-\hat{\boldsymbol{q}}{:}\nabla \boldsymbol{E} = -\sum_{i,j}^{x,y,z} \hat{q}_{ij}(\partial E_j/\partial i)$. The generalized quadrupole operator $\hat{\boldsymbol{q}}$ in Eq. (5) incorporates both electric quadrupole and magnetic dipole moments as the symmetric and asymmetric tensor components, respectively.[55] Although the generalized quadrupole in the following discussion and calculations takes account of both the electric quadrupole and magnetic dipole, the latter is negligibly small in the present tip condition.

Depending on whether dipole ($\hat{\boldsymbol{\mu}}$) or quadrupole ($\hat{\boldsymbol{q}}$) interactions contribute to the two excitation processes in the SFG process, four types of second-order dipolar polarization terms can arise: $\boldsymbol{P}_{\mu\mu}^{(2)}$, $\boldsymbol{P}_{q\mu}^{(2)}$, $\boldsymbol{P}_{\mu q}^{(2)}$, and $\boldsymbol{P}_{qq}^{(2)}$, as illustrated in Figure 4. The resulting SFG electric field $\boldsymbol{E}_{\mathrm{P,SFG}}$ at a position $\boldsymbol{r}$, generated by these dipolar polarizations located at the origin ($\boldsymbol{r}=\boldsymbol{0}$), is given by:

$$\boldsymbol{E}_{\mathrm{P,SFG}} = k^2 \frac{\exp(\mathrm{i}kr)}{r}(\boldsymbol{n} \times \boldsymbol{P}^{(2)}) \times \boldsymbol{n}, \tag{6}$$

where $k = \omega_{\mathrm{SFG}}/c$, $r = |\boldsymbol{r}|$, $\boldsymbol{n} = \boldsymbol{r}/r$, $\boldsymbol{P}^{(2)} = \boldsymbol{P}_{\mu\mu}^{(2)} + \boldsymbol{P}_{q\mu}^{(2)} + \boldsymbol{P}_{\mu q}^{(2)} + \boldsymbol{P}_{qq}^{(2)}$, and $c$ is the speed of light. In addition, the external electric field may also induce quadrupolar polarization $\boldsymbol{Q}^{(2)}$, leading to another four contributions: $\boldsymbol{Q}_{\mu\mu}^{(2)}$, $\boldsymbol{Q}_{q\mu}^{(2)}$, $\boldsymbol{Q}_{\mu q}^{(2)}$, and $\boldsymbol{Q}_{qq}^{(2)}$ (Figure 4). These quadrupolar polarizations also contribute to the SFG radiation as follows:

$$\boldsymbol{E}_{\mathrm{Q,SFG}} = -\mathrm{i}k^3 \frac{\exp(\mathrm{i}kr)}{r}\left(\boldsymbol{n} \times (\boldsymbol{Q}^{(2)} \cdot \boldsymbol{n})\right) \times \boldsymbol{n}. \tag{7}$$

Among these eight polarization components, only four terms $\boldsymbol{P}_{\mu\mu}^{(2)}$, $\boldsymbol{P}_{qq}^{(2)}$, $\boldsymbol{Q}_{q\mu}^{(2)}$, and $\boldsymbol{Q}_{\mu q}^{(2)}$ reflect information about the absolute up/down molecular orientation relative to surfaces, while other four terms are not orientation sensitive (see Supporting Information section 8 for details). Therefore, to accurately extract orientation information of interfacial molecules from the measured SFG spectra, it is essential to assess the relative contribution of each term. To this end, we performed electromagnetic field simulations to characterize the spatial distribution of the field $\mathbf{E}$ and its gradient $\nabla \mathbf{E}$ under the present tip condition and estimated SFG intensities arising from those dipolar and quadrupolar polarizations.

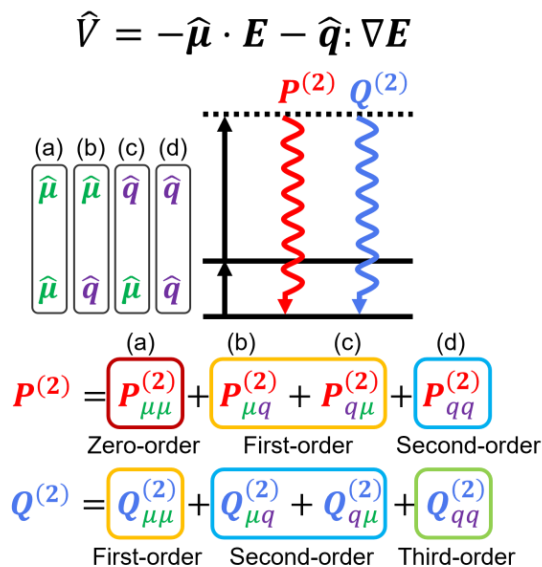

Figure 4. Schematic representation of SFG process involving dipolar and quadrupolar transitions. The light–molecule interaction ($\hat{V}$) involves the dipole–field coupling ($-\hat{\boldsymbol{\mu}} \cdot \mathbf{E}$) and the quadrupole–field-gradient coupling ($-\hat{\mathbf{q}}:\nabla \mathbf{E}$). In mechanism (a), both of two excitation processes in SFG are induced by dipole transition, while in mechanisms (b) and (c), one of them is replaced by a quadrupole transition. In mechanism (d), both excitation processes are induced by quadrupole transition. Since both dipolar ($\mathbf{P}^{(2)}$) and quadrupolar ($\mathbf{Q}^{(2)}$) polarizations are generated through second-order light–molecule interaction, eight polarizations are possible in total as shown in the two equations at the bottom of the figure. These polarization terms are classified according to the order of the quadrupole contribution involved.

Electromagnetic field simulations were conducted using the finite-difference time-domain (FDTD) method, based on the geometry shown in Figure 5a, to compute the spatial distribution of the electric field within the tip–substrate nanogap (Figure 5b). In these simulations, the tip apex radius was set to 50 nm, consistent with the actual size of the tip apex used in our experiments (Figure 1c). Here, we focus on the surface-normal component of the electric field ($E_Z$) because $E_Z$ is overwhelmingly enhanced compared to the surface-parallel components ($E_X$ and $E_Y$), as shown in Figure S7. As illustrated by the field cross sections along *X* and *Z* directions (Figure 5c and d, respectively), the electric field intensity varies over a spatial scale of approximately ~10 nm, which is significantly larger than the molecular size (~6 Å). Moreover, the experimentally confirmed TE-SFG spatial resolution of ~30 nm suggests the absence of an atomistic protrusion at the tip apex[75–77], effectively ruling out atomic/submolecular-scale strong field gradient within the nanogap. Therefore, the field electric gradient within the nanogap should be insignificant compared to the molecular dimensions, suggesting the quadrupole–field-gradient interactions play only a minor role under the present conditions.

To quantitatively verify this conclusion, we performed the quantum chemical calculation employing Qsac package[51] to obtain relevant susceptibility tensors and calculated TE-SFG intensities emitted from each polarization (see Supporting Information section 9 for the details of calculation procedures). The eight polarizations in Figure 4 can be classified into three groups according to the quadrupole order: (A) zero-quadrupole-order term $P^{(2)}_{\mu\mu}$, (B) first-quadrupole-order terms $P^{(2)}_{q\mu}$, $P^{(2)}_{\mu q}$, and $Q^{(2)}_{\mu\mu}$, and (C) higher-quadrupole-order terms $P^{(2)}_{qq}$, $Q^{(2)}_{\mu q}$, $Q^{(2)}_{q\mu}$, and $Q^{(2)}_{qq}$. Since the contribution of the higher-quadrupole-order terms (C) are clearly much smaller and negligible compared to that of the zero- and first-quadrupole-order terms, we focused our analysis on the zero- and first-order terms and calculated TE-SFG intensities arising from these contributions as a function of lateral position along the *X*-axis (Figure 5e). The calculated TE-SFG intensities indicate that the combined signal from $P^{(2)}_{q\mu}$ and $P^{(2)}_{\mu q}$ is approximately four orders of magnitude weaker than that from $P^{(2)}_{\mu\mu}$, and the signal from $Q^{(2)}_{\mu\mu}$ is seven orders of magnitude weaker ($P^{(2)}_{\mu\mu} \gg P^{(2)}_{q\mu} + P^{(2)}_{\mu q} \gg Q^{(2)}_{\mu\mu}$). These results clearly indicate that the TE-SFG emission from $Q^{(2)}_{\mu\mu}$ as well as $P^{(2)}_{q\mu}$ and $P^{(2)}_{\mu q}$ is negligible. Therefore, we can conclude that $P^{(2)}_{\mu\mu}$ term dominantly contributes to the observed TE-SFG signals under our experimental conditions, validating the applicability of the dipole approximation. Notably, the TE-SFG signal emitted by $P^{(2)}_{\mu\mu}$ arises from a ~10-nm-scale region beneath the tip (Figure 5e), consistent with the experimentally confirmed spatial resolution of ~30 nm (Figure 2c).

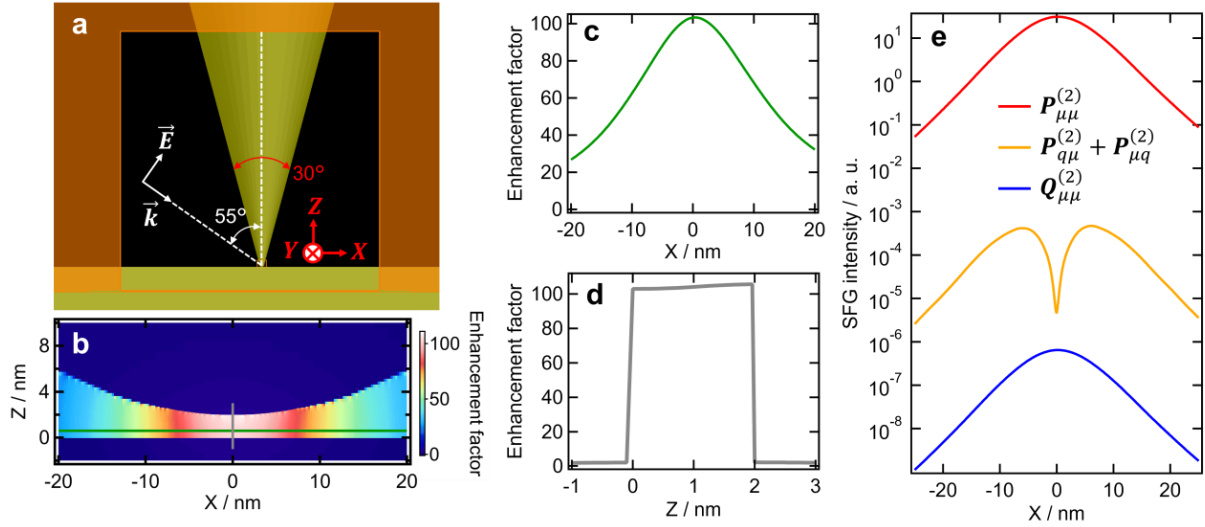

Figure 5. (a) The geometry of the gold tip, gold substrate, and incident light in the three-dimensional FDTD simulations. The opening angle of the tip and the incident angle of excitation light were 30° and 55°, respectively. The orange-coloured areas are allocated to the boundary condition of a perfectly matched layer. (b) Spatial distribution of $Z$-directed local field enhancement factor in $Z$–$X$ plane under 2-nm tip–substrate distance. The excitation wavelength is 1033 nm. (c, d) Plots of $|E_z|$ at 1033 nm (c) along the $X$-axis with $Y = 0$ nm and $Z = 0.6$ nm (the green horizontal line in b) and (d) along the $Z$-axis with $X = 0$ nm and $Y = 0$ nm (the grey vertical line in b). (e) Simulated SFG intensities arising from $\boldsymbol{P}^{(2)}_{\mu\mu}$ (red), $\boldsymbol{P}^{(2)}_{q\mu} + \boldsymbol{P}^{(2)}_{\mu q}$ (orange), and $\boldsymbol{Q}^{(2)}_{\mu\mu}$ (blue) induced within 4-MBT molecules on Au substrate. The SFG intensities were calculated for the methyl symmetric stretching mode of 4-MBT molecules and are plotted along the $X$-axis.

We then further discuss the physical origin of the magnitude relationship among the polarization terms: $\boldsymbol{P}^{(2)}_{\mu\mu} \gg \boldsymbol{P}^{(2)}_{q\mu} + \boldsymbol{P}^{(2)}_{\mu q} \gg \boldsymbol{Q}^{(2)}_{\mu\mu}$. The significant weakness of $\boldsymbol{Q}^{(2)}_{\mu\mu}$ can be qualitatively explained by the nature of quadrupolar radiation: it can be regarded as a superposition of radiation from two oppositely oriented dipoles, whose radiation mostly cancels out, resulting in significantly weaker net radiation compared to a single dipolar radiation. In contrast, the dominance of $\boldsymbol{P}^{(2)}_{\mu\mu}$ over $\boldsymbol{P}^{(2)}_{q\mu} + \boldsymbol{P}^{(2)}_{\mu q}$ requires more quantitative considerations. In the space-fixed coordinate system ($I, J, K, L \in \{X, Y, Z\}$), the $I$ components of these polarizations are expressed as follows:

$$P^{(2)}_{I,\mu\mu}(\omega_{\text{SFG}}) = \sum_{J,K}^{X,Y,Z} \chi^{D\mu\mu}_{IJK}(\omega_{\text{SFG}}; \omega_{\text{MIR}}, \omega_{\text{NIR}}) E_J(\omega_{\text{NIR}}) E_K(\omega_{\text{MIR}}) \quad (8)$$

$$P^{(2)}_{I,q\mu}(\omega_{\text{SFG}}) + P^{(2)}_{I,\mu q}(\omega_{\text{SFG}}) = \sum_{J,K,L}^{X,Y,Z} \chi^{Dq\mu}_{IJKL}(\omega_{\text{SFG}}; \omega_{\text{MIR}}, \omega_{\text{NIR}}) \frac{\partial E_J(\omega_{\text{NIR}})}{\partial L} E_K(\omega_{\text{MIR}})$$
$$+ \sum_{J,K,L}^{X,Y,Z} \chi^{D\mu q}_{IJKL}(\omega_{\text{SFG}}; \omega_{\text{MIR}}, \omega_{\text{NIR}}) E_J(\omega_{\text{NIR}}) \frac{\partial E_K(\omega_{\text{MIR}})}{\partial L} \quad (9)$$

where $\chi^{D\mu\mu}_{IJK}$, $\chi^{Dq\mu}_{IJKL}$, and $\chi^{D\mu q}_{IJKL}$ are the second-order nonlinear susceptibility tensors in the space-fixed coordinate. The derivation of these equations is given in Supporting Information section 8. In eq. (8), the contribution of $P^{(2)}_{I,\mu\mu}$ arises from the $\chi^{D\mu\mu}_{ZZZ} E_Z(\omega_{\text{NIR}}) E_Z(\omega_{\text{MIR}})$ term ($(I,J,K) = (Z,Z,Z)$), because the surface-normal electric field component ($E_Z$) is overwhelmingly enhanced under the tip compared to in-plane components (Figure S7e). Meanwhile, the dominant term in eq. (9) for $P^{(2)}_{I,q\mu} + P^{(2)}_{I,\mu q}$ is $\chi^{Dq\mu}_{XXZZ}(\partial E_X(\omega_{\text{NIR}})/\partial Z) E_Z(\omega_{\text{MIR}}) \approx \chi^{Dq\mu}_{XXZZ}(\partial E_Z(\omega_{\text{NIR}})/\partial X) E_Z(\omega_{\text{MIR}})$ term ($(I,J,K,L) = (X,X,Z,Z)$), because the field gradients of $\partial E_X(\omega_{\text{NIR}})/\partial Z$ and $\partial E_Z(\omega_{\text{NIR}})/\partial X$ are predominant compared to other field gradient components in the present near-field condition. The field gradient is nearly symmetric (e.g. $\partial E_Z/\partial X = \partial E_X/\partial Z$), implying that the contribution of magnetic fields and magnetic dipolar transitions are relatively unimportant compared to the enhanced electric field. Comparing the dominant terms of eqs. (8) and (9), we confirmed the following relationship:

$$\left| \chi^{D\mu\mu}_{ZZZ} E_Z(\omega_{\text{NIR}}) E_Z(\omega_{\text{MIR}}) \right| \gg \left| \chi^{Dq\mu}_{XXZZ} \frac{\partial E_X(\omega_{\text{NIR}})}{\partial Z} E_Z(\omega_{\text{MIR}}) \right| \quad (10)$$

This relation warrants the dominance of the dipolar contribution over quadrupolar one under the present tip conditions. To quantitatively understand this relationship, we evaluated the ratio of the left and right hand sides of eq. (10), which is given by the product of $\chi^{Dq\mu}/\chi^{D\mu\mu}$ and $(\partial E_X(\omega_{\text{NIR}})/\partial Z)/E_Z(\omega_{\text{NIR}})$. The ratio $\chi^{Dq\mu}/\chi^{D\mu\mu}$ reflects the molecular length scale (typically a few Å- to nm-scale),[51,78] while $(\partial E_X(\omega_{\text{NIR}})/\partial Z)/E_Z(\omega_{\text{NIR}}) \approx (\partial E_Z(\omega_{\text{NIR}})/\partial X)/E_Z(\omega_{\text{NIR}})$ scales with the inverse of the curvature radius of the tip apex[75] (~50 nm in our case, Figure 1c). These values yield an estimate of $6\ \text{Å} \times (50\ \text{nm})^{-1} \sim 10^{-2}$ for the ratio of the left and right hand sides of eq. (10), indicating that the quadrupolar term is approximately two orders magnitude smaller than the dipolar term. Accordingly, the relative quadrupole-to-dipole TE-SFG intensity ratio is estimated to be $(10^{-2})^2 \sim 10^{-4}$, consistent with the results shown in Figure 5e. This analysis supports the conclusion that the electric field gradient beneath the tip[73,74] is too small to induce substantial quadrupole contributions, thereby validating the dipole approximation for interpreting the TE-SFG signals under the present tip conditions.

Note that future studies employing subnanometer-scale tip sharpness or atomistic protrusions to achieve higher spatial resolution may generate stronger electric field gradients on spatial scales comparable to or even smaller than molecular dimensions. In such case, the ratio $(\partial E_X(\omega_{\text{NIR}})/\partial Z)/E_Z(\omega_{\text{NIR}}) \approx (\partial E_Z(\omega_{\text{NIR}})/\partial X)/E_Z(\omega_{\text{NIR}})$ is expected to approach $\text{Å}^{-1}$-scale, resulting in a comparable magnitude of left and right sides of eq. (10). This implies that quadrupole-induced SFG radiation, particularly from $\boldsymbol{P}_{q\mu}^{(2)} + \boldsymbol{P}_{\mu q}^{(2)}$, can no longer be neglected. Moreover, the contribution of $\boldsymbol{P}_{qq}^{(2)}$, which involves double quadrupole–filed-gradient interactions (mechanism (d) in Figure 4) may also become significant. Therefore, as TE-SFG evolves toward molecular/atomic-scale resolution, such quadrupolar contributions would become essential, as reported in previous tip-enhanced Raman spectroscopy.[73,74] The theoretical framework presented here provides a basis for rigorously addressing these effects in TE-SFG nanoscopy.

**Domain-specific adsorption structures**

Given that the dipole approximation has been validated under the present experimental conditions, the sign of $\text{Im}\left(\chi_{\text{R},a}^{(2)}\right)$ reliably reflects the absolute up/down molecular orientation.[1–3,5–9] This orientation sensitivity represents an advantage of the second-order TE-SFG over linear and odd-order nonlinear near-field nanospectroscopies, which are typically less sensitive to interfacial polarity. The physical origin of this absolute orientation sensitivity can be attributed to two key factors that determine the sign of $\text{Im}\left(\chi_{\text{R},a}^{(2)}\right)$: (1) the directional cosines of the molecular coordinates relative to the surface normal and (2) the molecular hyperpolarizability $\beta_{ijk}$ associated with each vibrational mode defined by the product of the transition Raman polarizability tensor $(\partial \alpha_{ij}/\partial u_a)$ and the dipole moment $(\partial \mu_k/\partial u_a)$, where $u_a$ is the normal coordinate of the vibrational mode $a$. More rigorous expression for the factor determining the sign of $\text{Im}\left(\chi_{\text{R},a}^{(2)}\right)$ is provided in eq. S52 in Supporting Information section 9. Quantum chemical calculations for the 4-MBT molecule indicate that $\beta_{ijk}$ is negative for all experimentally observed vibrational modes (Supporting Information section 9). In this case, the negative $\text{Im}(\chi_{R,a}^{(2)})$ signals obtained through TE-SFG (Figure 3c and d) and FF-SFG (Figure S3b) measurements correspond to positive directional cosines of the methyl groups, indicating that the hydrogen atoms of the methyl group are oriented upward relative to the surface (H-up configuration), consistent with the known absolute orientation of 4-MBT molecules.[57,58] These results demonstrate that our TE-SFG[50] accurately captures local up/down molecular orientations on the nanoscale. This is in stark contrast to FF-SFG, which provides macroscopically averaged orientation information. Thus, the present work extends the absolute orientation sensitivity of the second-order nonlinear SFG spectroscopy[1–3,5–9] into the near-field-based spatially resolved measurements beyond the diffraction limit.

In addition to the absolute up/down molecular orientation, the present TE-SFG approach also allows us to extract more detailed structural parameters of molecular adsorption geometry: the molecular tilt angle $\theta$, defined as the angle between the trigonal axis of the methyl group and the surface normal (Figure 6). Since the transition dipole moment of the $r^+$ mode aligns with the trigonal axis of methyl group, whereas that of the $r^-$ mode is oriented perpendicularly to that of the $r^+$ mode, the SFG intensity ratio of these bands ($A_{r^-}/A_{r^+}$) generally increases monotonically with $\theta$.[66,68] By numerically calculating the relative SFG intensities as a function of $\theta$ (Supporting Information section 10), we obtained quantitative calibration curves corelating $A_{r^-}/A_{r^+}$ ratio with $\theta$ (Figure 6). Comparison of the experimental SFG intensities with these curves yielded estimated tilt angles of 19° and 30° for the major and minor domains, respectively (Figure 6). In addition, similar monotonical $\theta$ dependences were obtained for other intensity ratios involving $r_{ph}$ bands: $A_{ph}/A_{r^+}$ (Figure S12a) and $A_{r^-}/A_{ph}$ (Figure S12b), from which $\theta$ values were independently derived.

By averaging the estimates from Figure 6 and Figure S12, the tilt angles $\theta$ for the major and minor domains were determined to be $(20 \pm 2)°$ and $(33 \pm 8)°$, respectively (Table S8), where the uncertainties represent standard deviations across three independent estimation methods (Figures 6 and S12). These results indicate that the molecules in the minor domain adopt a more inclined adsorption configuration than those in the major domain. Notably, these tilt angles are consistent with previous density functional theory calculations for benzenethiol molecules on Au(111) surface, which predicted distinct SAM aggregated structures with tilt angles of 21° and 33°.[79] Despite the difference in molecular species, the structural similarity between 4-MBT and benzenethiol molecules dominated by a phenyl ring supports the relevance of this comparison. While determining molecular tilt angles is often challenging even with high-resolution STM imaging,[80] our TE-SFG technique enables their direct extraction from vibrational spectra. This highlights the capability of TE-SFG nanoscopy to probe nanoscale structural heterogeneity within interfacial molecular aggregates.

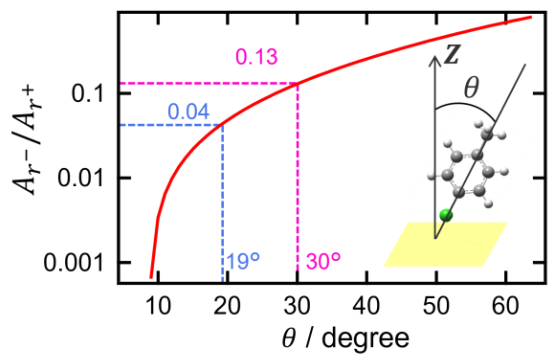

Figure 6. Calculated $\theta$ dependence of the relative TE-SFG intensity ratio between the symmetric ($r^+$) and asymmetric ($r^-$) stretching modes of the methyl group. The blue and magenta horizontal broken lines represent the experimentally obtained intensity ratios for the major and minor domains, respectively, and the corresponding vertical broken lines indicate the extracted tilt angles. Inset: schematic illustration of the molecular tilt angle $\theta$, defined as the angle between the trigonal axis of the methyl group and the surface normal (Z-axis).

**Conclusion**

We have achieved the diffraction-unlimited vibrational SFG nanoscopy by developing the near-field TE-SFG scheme with a spatial resolution down to ~10 nm, a nearly two-orders-of-magnitude improvement over micrometer-scale diffraction-limited resolution of conventional far-field SFG. To rigorously interpret the observed domain-dependent TE-SFG spectra, we developed a comprehensive theoretical framework describing microscopic mechanisms of the TE-SFG process involving both dipole–field interactions and higher-order multipole effects. Our combined experimental and theoretical analysis revealed that TE-SFG spectra faithfully reflect not only nanoscale structural variations but also absolute up/down molecular orientations within nanometric interfacial molecular domains, providing direct access to site-specific molecular information beyond the reach of conventional far-field SFG and STM imaging. Given that these demonstrations were realized under ambient conditions, TE-SFG nanoscopy offers a powerful platform for investigating interfacial molecular systems in realistic environments. These results collectively position TE-SFG as a transformative vibratioal spectroscopy for resolving nanoscale structural heterogeneity and molecular dynamics at practical interfaces where absolute orientation plays key roles.[3,6,13,81–87]

ASSOCIATED CONTENT

**Supporting Information**.

The following files are available free of charge:

Experimental details, additional STM images, far-field SFG spectrum, optimized parameters for spectral fitting analyses, assignments for the observed vibrational bands, field enhancement mechanisms governing TE-SFG process, spatial distribution of electric fields, theroretical backgrounds of SFG, evaluation of dipolar and quadrupolar contributions in the TE-SFG signals, deviations of the relation between molecular tilt angles and SFG intensity, and discussion on the validity of the numerical calculation.

AUTHOR INFORMATION

**Corresponding Authors**

Atsunori Sakurai — Department of Materials Molecular Science, Institute for Molecular Science (IMS), Okazaki, Aichi 444-8585, Japan; The Graduate University for Advanced Studies, SOKENDAI, Okazaki, Aichi, 444-8585, Japan; Email: asakurai@ims.ac.jp

Akihiro Morita — Department of Chemistry, Graduate School of Science, Tohoku University, Sendai, Miyagi 980-8578, Japan; Email: morita@tohoku.ac.jp

Toshiki Sugimoto — Department of Materials Molecular Science, Institute for Molecular Science (IMS), Okazaki, Aichi 444-8585, Japan; The Graduate University for Advanced Studies, SOKENDAI, Okazaki, Aichi 444-8585, Japan; Email: toshiki-sugimoto@ims.ac.jp

**Authors**

Shota Takahashi — Department of Materials Molecular Science, Institute for Molecular Science (IMS), Okazaki, Aichi 444-8585, Japan

Koichi Kumagai — Department of Chemistry, Graduate School of Science, Tohoku University, Sendai, Miyagi 980-8578, Japan

Tatsuto Mochizuki — Department of Materials Molecular Science, Institute for Molecular Science (IMS), Okazaki, Aichi 444-8585, Japan; The Graduate University for Advanced Studies, SOKENDAI, Okazaki, Aichi 444-8585, Japan

Tomonori Hirano — Department of Chemistry, Graduate School of Science, Tohoku University, Sendai, Miyagi 980-8578, Japan.

**Notes**

The authors declare no competing financial interest.

ACKNOWLEDGMENT

We thank M. Aoyama, T. Kondo, N. Mizutani, T. Kikuchi, and T. Toyoda at the Equipment Development Center, Institute for Molecular Science (IMS), and E. Nakamura at the UVSOR synchrotron facility of IMS for their technical assistance. Scanning electron microscopy observation of tips was conducted at the Institute for Molecular Science, supported by "Advanced Research Infrastructure for Materials and Nanotechnology in Japan (ARIM)" of the Ministry of Education, Culture, Sports, Science and Technology (MEXT), Proposal Number JPMXP1223MS5022. T.S. acknowledges financial support from JSPS KAKENHI Grant-in-Aid for Scientific Research (A) (19H00865), Grant-in-Aid for Transformative Research Areas (A) (24H02205); JST-PRESTO (JPMJPR1907); JST-FOREST (JPMJFR221U); and JST-CREST (JPMJCR22L2). A.M. acknowledges financial support from JSPS KAKENHI Grant-in-Aid for Scientific Research (A) (20H00368); and for Scientific Research (B) (25K01721). A.S. acknowledges financial support from JSPS KAKENHI Grant-in-Aid for Scientific Research (B) (23H01855); for Early-Career Scientists (20K15236); Casio Science Promotion Foundation (38-06); and Research Foundation for Opto-Science and Technology. S.T. acknowledges financial support from Grant-in-Aid for JSPS Fellows (22KJ3099).

# Supporting Information: Diffraction-Unlimited Tip-Enhanced Sum-Frequency Vibrational Nanoscopy

*Shota Takahashi[1], Koichi Kumagai[2], Atsunori Sakurai[1,3]\*, Tatsuto Mochizuki[1,3], Tomonori Hirano[2], Akihiro Morita[2]\*, and Toshiki Sugimoto[1,3]\**

[1]Department of Materials Molecular Science, Institute for Molecular Science (IMS), Okazaki, Aichi 444-8585, Japan
[2]Department of Chemistry, Graduate School of Science, Tohoku University, Sendai, Miyagi 980-8578, Japan.
[3]The Graduate University for Advanced Studies, SOKENDAI, Okazaki, Aichi, 444-8585, Japan

\*Corresponding authors
E-mail: asakurai@ims.ac.jp, morita@tohoku.ac.jp, toshiki-sugimoto@ims.ac.jp

CONTENTS

# 1. Experimental procedures

The experimental procedures for the preparation of a STM tip, formation of the self-assembled monolayer (SAM), and TE-SFG measurements are described elsewhere.[1] Briefly, gold tips for STM were prepared by electrochemical etching of gold wires and observed using a field emission scanning electron microscopy system (SU-6600) to investigate the tip apex structure. The accelerating voltage and working distance were 5 kV and 4 mm, respectively.

For the substrate to form a model molecular monolayer, a 200-nm-thick gold thin film vapor-deposited on a mica substrate (UNISOKU) was employed. The substrate was first annealed in a butane flame to achieve an atomically flat Au(111) surface and cooled down to room temperature. Then, the Au substrate was immersed in 1 mM ethanolic solution of 4-Methylbenzenethiol (4-MBT) at room temperature for 24 hours to form a ~6 Å thick self-assembled monolayer (SAM) of 4-MBT on its surface.[2] Thereafter, the Au substrate was picked up from the solution, rinsed with pure ethanol, and dried with air just before being placed into our STM chamber. Note that 4-MBT has a relatively short molecular length (~6 Å) compared to typical straight chain alkanethiols,[3] resulting in a smaller plasmonic nanocavity gap between the Au tip and substrate,[1] which is advantageous to obtain stronger near-field TE-SFG signals.

All the TE-SFG measurements were performed at room temperature and ambient pressure conditions. Our optical system is based on an amplified Yb-fiber laser (1033 nm, 280 fs, 40 W, 50 MHz; Monaco-1035-40-40, Coherent). The fundamental output from the laser was divided into two portions by a beam splitter. The first portion (120 nJ) was used to drive a commercially available optical parametric oscillator (OPO, Levante IR, APE), generating two kinds of wavelength-tunable IR pulses: signal (1.3 – 2 μm) and idler (2.1 – 5 μm). The second portion was passed through an ultranarrow bandpass filter to narrow down the spectral width into ~10 cm$^{-1}$ (Figure S1a). We focused the spatially and temporally overlapped idler pulses ($\omega_{MIR}$: scan range 2800–3050 cm$^{-1}$, 300 fs, FWHM: 70 cm$^{-1}$, 0.4 nJ) and the narrow-band fundamental wave ($\omega_{NIR}$: 9674.0 cm$^{-1}$ (1033.7 nm), 1 ps, FWHM: 10 cm$^{-1}$, 10 pJ) onto the tip–sample nanogap to induce TE-SFG process at ~800 nm. Both pulses were $p$-polarized. The average (peak) power densities of the mid- and near-IR lasers at the focal point were $5.0 \times 10^3$ ($3.3 \times 10^8$) W/cm$^2$ and $1.5 \times 10^3$ ($2.9 \times 10^7$) W/cm$^2$, respectively. The STM sample bias and tunneling current setpoint were set to 0.1 V and 100 pA, respectively. This bias voltage is sufficiently low to suppress the contributions from DC-field-induced third-order SFG[4–6] across the nanogap.

The far-field SFG (FF-SFG) measurements were also performed by maintaining the tip–sample distance at more than 1 μm to prevent the excitation of the gap-mode plasmon. During the FF-SFG measurements, the average (peak) power densities of the mid- and near-IR lasers at the focal point were set to $5.0 \times 10^3$ ($3.3 \times 10^8$) W/cm$^2$ and $1.5 \times 10^4$ ($2.9 \times 10^8$) W/cm$^2$, respectively.

Typical spectral width of the idler lasers is approximately 70 cm$^{-1}$, which is insufficient to cover the whole frequency region of C-H stretching modes. Thus, we repeatedly measured SFG intensity spectra while tuning the central wavenumber of the mid-IR lasers across the C-H stretching region of a 4-MBT molecule. This experimental scheme allowed us to obtain the overall vibrational spectra in C-H stretching region covering both methyl stretching (2800–3000 cm$^{-1}$) and phenyl C–H stretching (3000–3050 cm$^{-1}$) modes. A series of mid-IR spectra used in our experiments are shown in Figure S1b.

Comparison of mid-IR idler output centered at 2918 cm$^{-1}$ (the red curve in Figure S1b) and typical TE-SFG intensity spectrum obtained for bare Au substrate (the red broken curve in Figure 1f in the main text) is shown in Figure S1c. The slightly broadened feature of TE-SFG spectra results from the spectral convolution of the mid-IR pulse with a narrowband 1033-nm pulse shown in Figure S1a. The spectral matching indicates that the shape of TE-SFG spectra is determined by the spectral distribution of these two excitation lights.

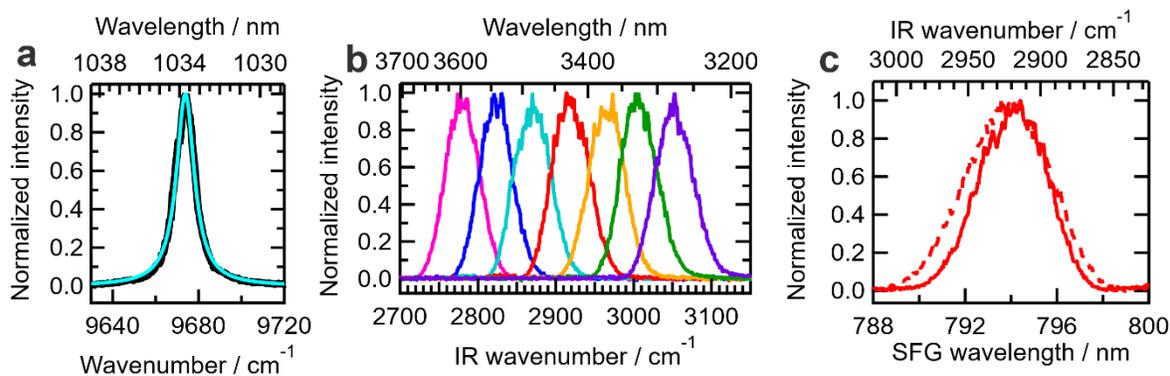

Figure S1. (a) Spectrum of the output from the Yb-fiber laser transmitted after the ultranarrow bandpass filter (black). Light blue curve is the result of curve fitting with Lorentz function. The central wavenumber (wavelength) was 9674.0 cm$^{-1}$ (1033.7 nm), and the full width at half maximum (FWHM) was 10.1 cm$^{-1}$. (b) Spectra of the mid-infrared (IR) idler outputs from the optical parametric oscillator (OPO) pumped by 1030 nm. The TE-SFG spectra shown in Fig. 1 and 2 in the main text were obtained by using 2918-cm$^{-1}$-centered pulse (red). Similar TE-SFG measurements were repeatedly performed by changing the center wavenumber of idler pulses, and the individual results were connected to produce the overall TE-SFG spectra shown in Figure 3a–c in the main text. (c) Comparison of mid-IR idler output centered at 2918 cm$^{-1}$ (solid curve, the same as the red curve in Figure S1b) and TE-SFG intensity spectrum obtained for bare Au substrate (broken curve, the same as the red broken curve in Figure 1e in the main text). Note that the mid-IR wavenumber is displayed in the top axis, and the corresponding values of the top axis are converted from the mid-IR wavenumber to the SFG wavelength and displayed in the bottom axis.

## 2. STM images of 4-MBT SAM on a Au substrate

In this section, we focus on the STM images of 4-MBT SAM on a Au substrate and provide the characterization of "the major domain" and "the minor domain", which were introduced in Figure 2a in the main text. Figure S2a shows a STM image of the 4-MBT-adsorbed Au substrate measured over 778 nm × 778 nm region. Observation of the Au(111) monoatomic steps indicates that the surface is atomically flat. Moreover, a number of island-like structures (adatom islands), which are typical of SAMs of arenethiols on Au(111),[7,8] confirm the formation of a single monolayer of 4-MBT. Notably, closely looking at the image, we can find a local region with fewer adatom islands present compared to surrounding regions (black round rectangle in Figure S2a). The difference in the island density becomes more clearly visible in a magnified view of the boundary region between the island-rich and island-less regions (Figure S2b). Since the island-rich region occupies predominantly larger areas on the surface, we denoted this island-rich region as "major domain" and the island-less region as "minor domain".

Furthermore, by comparing the STM images acquired with (Figure S2c) and without (Figure S2b) excitation laser irradiation, we can ensure that the excitation laser irradiation does not affect or damage the structure of the tip apex during the TE-SFG measurements. Although laser irradiation seems to slightly increase the noise level on the STM image, angstrom-scale Au monoatomic step structures were clearly captured as shown in Figure S2c. Moreover, the observed step heights (Figure S2d and e) are consistent with previously reported values of Au monoatomic step heights[9–11]. Therefore, we can conclude that the tip structure did not evolve during the tip-enhanced nonlinear optical measurements.

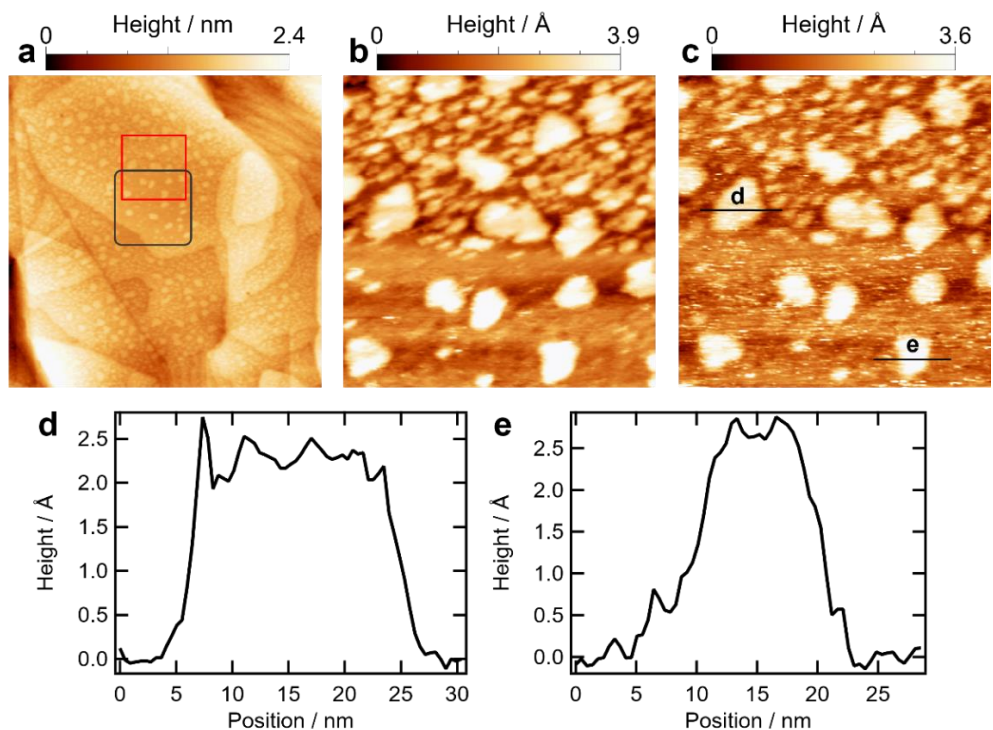

Figure S2. (a–c) STM images of 4-MBT SAM acquired at a sample bias of 0.1 V and a tunneling current setpoint of 100 pA. The sizes of measured regions are 778 nm × 778 nm (a) and 155 nm × 155 nm (b and c). The protrusions that can be seen in (a–c) are derived from Au adatom islands. The black round rectangle in (a) represents the domain with fewer islands compared to the surrounding region. The images (b) and (c) are magnified views of the red squared region in (a), acquired without and with irradiating the STM nanogap by the excitation lasers for TE-SFG experiments, respectively. Note that the image (c) is identical to that shown in Figure 2a in the main text. (d, e) Height profiles for the solid black lines labeled "d" and "e" in (c).

## 3. Far-field SFG spectrum

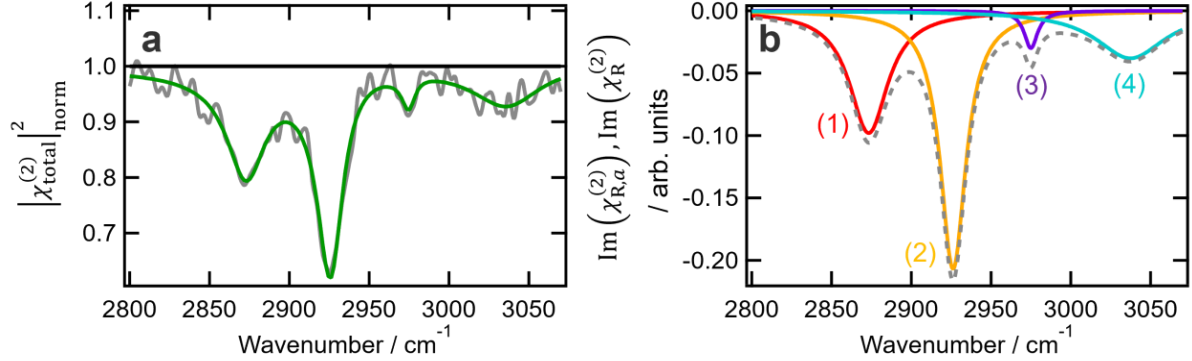

Figure S3. (a) $|\chi^{(2)}_{total}|^2_{norm}$ spectra in the CH stretching region of 4-MBT molecules obtained by far-field SFG measurements. Green curves are the results of curve fitting based on eq. (4) in the main text. The pulse energies of the mid-IR and near-IR excitation light were 400 pJ and 100 pJ, respectively. (b) Spectra of $\text{Im}(\chi^{(2)}_{R,a})$ corresponding to four different vibrational modes of a 4-MBT molecule ($a = 1,2,3,4$). Gray broken curve is the $\text{Im}(\chi^{(2)}_R)$ spectra obtained by summing the four $\text{Im}(\chi^{(2)}_{R,a})$ spectra.

## 4. Optimized parameters for spectral fitting analyses

Table S1. Optimized fitting parameters used in the fitting analyses for the TE-SFG spectra shown in Figure 3 in the main text and the FF-SFG spectrum shown in Figure S3a. The fitting was performed under the constraint that the FWHMs of $\text{Im}(\chi^{(2)}_{R,a})(= 2\Gamma_a)$ be larger than that of the near-IR excitation pulse (10.1 cm$^{-1}$, see Figure S1a).

| Parameters | TE-SFG from major domain (Figure 3a) | TE-SFG from minor domain (Figure 3b) | FF-SFG (Figure S3a) |
|---|---|---|---|
| $\phi^{(2)}_{NR}$ / degree | 34 ± 1 | 90 ± 3 | 84 ± 1 |
| $A'_1$ / arb. units | −1.62 ± 0.08 | −0.46 ± 0.1 | −1.44 ± 0.07 |
| $\omega_1$ / cm$^{-1}$ | 2866.3 ± 0.4 | 2856.9 ± 1.6 | 2873.3 ± 0.5 |
| $\Gamma_1$ / cm$^{-1}$ | 9.2 ± 0.6 | 12.3 ± 3.0 | 14.6 ± 0.8 |
| $A'_2$ / arb. units | −3.79 ± 0.08 | −3.4 ± 0.2 | −2.02 ± 0.04 |
| $\omega_2$ / cm$^{-1}$ | 2918.4 ± 0.2 | 2924.5 ± 0.8 | 2926.2 ± 0.2 |
| $\Gamma_2$ / cm$^{-1}$ | 10.1 ± 0.3 | 19.0 ± 1.2 | 9.7 ± 0.3 |
| $A'_3$ / arb. units | −0.23 ± 0.06 | −0.50 ± 0.2 | −0.15 ± 0.03 |
| $\omega_3$ / cm$^{-1}$ | 2967.9 ± 1.6 | 2957.8 ± 1.6 | 2975.0 ± 0.7 |
| $\Gamma_3$ / cm$^{-1}$ | 7.6 ± 2.5 | 11.3 ± 3.1 | 5.1 ± 1.1 |
| $A'_4$ / arb. units | −1.41 ± 0.08 | −1.6 ± 0.1 | −1.01 ± 0.06 |
| $\omega_4$ / cm$^{-1}$ | 3022.4 ± 0.5 | 3035.2 ± 1.3 | 3037.4 ± 1.4 |
| $\Gamma_4$ / cm$^{-1}$ | 10.6 ± 0.7 | 22.4 ± 2.0 | 26.5 ± 2.3 |

The phase of the vibrationally non-resonant susceptibility ($\phi_{NR}^{(2)}$) is generally related to the surface electronic properties of metal substrates, including electronic density of states and band structures.[12] Thus, the different $\phi_{NR}^{(2)}$ values obtained for TE-SFG from the major and minor domains may be attributed to distinct surface electronic structures across the two domains, possibly associated with variations in the density of the adatom island structures (Figure 2a in the main text). More detailed understanding of the physical origin of the different $\phi_{NR}^{(2)}$ values requires comprehensive investigation of domain-dependent local electronic structures of the Au substrate and their influence on the near-field nonlinear optical responses, which lies beyond the scope of the present study.

## 5. Assignments for vibrationally resonant TE-SFG signals

In this section, we provide detailed discussion on the assignments for peaks (1)–(4) in TE-SFG spectra shown in Figure 3c and d in the main text and Figure S3b. To theoretically investigate the vibrational modes of a 4-MBT molecule, we performed structural optimization and normal mode analysis through density functional theory calculations using the B3LYP hybrid functional[13,14] in Gaussian 16 package.[15] Given that 4-MBT molecules form covalent bonds with Au atoms in the 4-MBT SAM sample, we included Au atoms to form a 4-MBT-Au$_3$ cluster, thereby accounting for the influence of Au atoms. In the calculation, aug-cc-pVTZ basis sets[16] were used for the C, H, S atoms while Au atoms were described with LanL2DZ basis set and effective core potentials.[17]

As shown in Figure S4, multiple vibrational modes relevant to the experimentally observed TE-SFG spectra were identified: $r_{d1}$ and $r_{d2}$: methyl deformation modes (Figure S4a and b); $r^+$: methyl symmetric stretching mode (Figure S4c); $r_1^-$ and $r_2^-$: methyl asymmetric stretching modes (Figure S4d and e); $r_{ph1}$ and $r_{ph2}$: C–H stretching modes in the benzene ring (Figure S4f and g).

As discussed in previous studies,[18,19] experimentally observed peaks (1) and (2) are attributed to a doublet arising from the Fermi resonance effect between $r^+$ mode and the overtone of $r_{d1}$ and $r_{d2}$ modes. The peak (3) originates from $r_1^-$ and $r_2^-$ modes. In molecules with high symmetry such as CH$_3$I ($C_{3v}$), the asymmetric stretching modes of the methyl group are degenerate.[20] In contrast, such symmetry is broken in a 4-MBT molecule and the degeneracy is lifted, resulting in peak splitting as seen in peaks d and e in Figure S4h and i. Nonetheless, the experimentally obtained $\chi_R^{(2)}$ spectra exhibited a single peak in the relevant frequency region,[21] which resembles the case of $C_{3v}$ symmetry. This mismatch between the experiments and calculations can be understood by considering the rotational motion of methyl groups. The rotational barrier of the methyl group is merely ~5 cm$^{-1}$,[22] suggesting that under the experimental conditions at room temperature (~200 cm$^{-1}$), the methyl group undergoes nearly free rotation. Therefore, the potential experienced by the methyl groups can be regarded as almost cylindrically symmetric under the actual experimental conditions, and thus the energy levels of the $r_1^-$ and $r_2^-$ should be effectively degenerate, resulting in a single peak for the methyl asymmetric vibrations.

Note that since lowering temperature should suppress the rotational motion of methyl groups, experiments under cryogenic temperature may capture two split $r_1^-$ and $r_2^-$ bands.[23] The remaining $r_{ph1}$ and $r_{ph2}$ modes nearly degenerate and are attributed to peak (4) in the TE-SFG spectra. The experimentally determined center frequencies of peaks (1)–(4) and assignments for those peaks are summarized in Table S2.

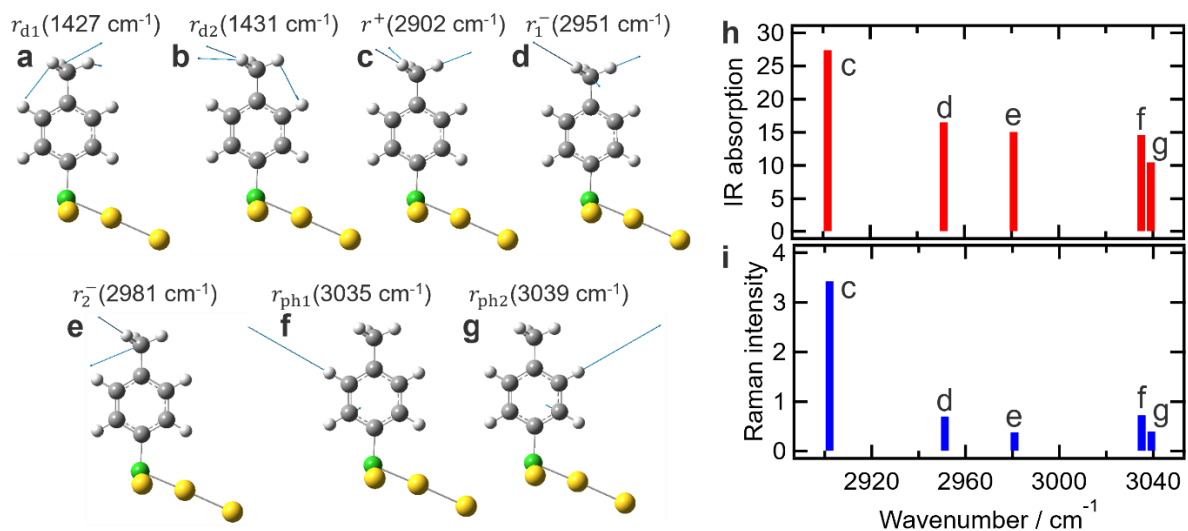

Figure S4. (a–g) The vibrational modes of a 4-MBT-Au$_3$ cluster. The white, gray, green, and yellow spheres represent hydrogen, carbon, sulfur, and gold atoms, respectively. In the main text, ($r_1^-$, $r_2^-$) modes and ($r_{ph1}$, $r_{ph2}$) modes are collectively referred to as $r^-$ and $r_{ph}$, respectively. The values in the parentheses represent the calculated vibrational frequencies of each mode. To correct for anharmonicity effects, the wavenumber values were scaled by a factor of 0.959. All structures were visualized using GaussView 6.[24] (h, i) Calculated IR absorption (h) and Raman intensity spectra (i) in the 2890–3050 cm$^{-1}$ range. Peaks corresponding to the vibrational modes shown in (c–g) are indicated.

Table S2. TE-SFG band assignments. The first three columns are peak frequencies obtained from TE-SFG experiments for the major and minor domains and FF-SFG experiments. The frequency values were adopted from Table S1. The numbers in the parentheses represent the corresponding vibrational bands indicated in Figures 3c and d in the main text. The next column shows calculated vibrational frequencies of a 4-MBT-Au$_3$ cluster (Figure S4). To correct for anharmonicity effects, the calculated wavenumber values were scaled by a factor of 0.959. Band assignments shown in the last column are based on the comparison between experimental and simulated values and a previous report on 4-MBT SAM.[19]

| Experiments | | | Simulations for a 4-MBT-Au$_3$ cluster | Assignments |
|---|---|---|---|---|
| TE-SFG Major domain | TE-SFG Minor domain | FF-SFG | | |
| Frequencies / cm$^{-1}$ | Frequencies / cm$^{-1}$ | Frequencies / cm$^{-1}$ | Frequencies / cm$^{-1}$ | |
| (1) 2866.3 ± 0.4 (2) 2918.4 ± 0.2 | (1) 2856.9 ± 1.5 (2) 2924.5 ± 0.9 | (1) 2873.0 ± 0.5 (2) 2925.7 ± 0.2 | 1426.75×2 1430.71×2 2901.98 | Fermi resonance formed by $r_{d1}$, $r_{d2}$, and $r^+$ modes ($r^+_{FR1}, r^+_{FR2}$) |
| (3) 2967.9 ± 1.6 | (3) 2957.8 ± 1.6 | (3) 2974.8 ± 0.4 | 2951.19 2980.84 | $r^-_1, r^-_2$ |
| (4) 3022.4 ± 0.5 | (4) 3035.2 ± 1.3 | (4) 3035.1 ± 1.3 | 3035.00 3039.17 | $r_{ph1}, r_{ph2}$ |

**6. Field enhancement mechanisms governing TE-SFG process—Broadband optical response spanning infrared to visible region—**

    **6.1   Comprehensive expression of TE-SFG intensity**

As discussed in the main text, we observed vibrationally resonant and non-resonant TE-SFG signals. In these TE-SFG emission processes, the optical wavelength range over which tip enhancement effectively works is a critical factor that determines the overall efficiency of TE-SFG. Since TE-SFG involves mid-IR-to-visible drastic frequency conversion between the incoming and outgoing light, a spectrally broad plasmonic enhancement that can simultaneously affect such separated frequencies is critical for realizing efficient generation of TE-SFG.[1,25] In the main text, however, we omitted the detailed description of the spectral properties of field enhancement strength in order to maintain clarity and conciseness. Instead, we focused on demonstrating the nanoscale spatial resolution of TE-SFG and on analyzing and interpreting the molecular vibrationally resonant TE-SFG signals. Thus, in this section, we provide an overview of the field enhancement mechanisms governing TE-SFG process by explicitly incorporating the plasmonic filed enhancement into the discussion.[1,25]

SFG involves two-photon excitation (generation of nonlinear polarization $P^{(2)}$) and one-photon emission (radiation from $P^{(2)}$) processes. The excitation and radiation processes are simultaneously amplified at the angstrom-scale gap in STM through plasmonic field enhancement, leading to a substantial increase in SFG signals. Consequently, the overall TE-SFG intensity ($I_{\text{TESFG}}$) can be described by using the incident field enhancement factor ($K_{\text{gap}} \equiv E_{\text{gap}}/E_0$), the enhanced emission efficiency from $P^{(2)}$ ($L_{\text{gap}}$), and second-order nonlinear optical susceptibility ($\chi^{(2)}$) as follows:[25]

$$I_{\text{TESFG}}(\omega_{\text{SFG}}) \propto \left|L_{\text{gap}}(\omega_{\text{SFG}})\right|^2 \left|\chi^{(2)}\right|^2 \left|K_{\text{gap}}(\omega_{\text{NIR}})E_0(\omega_{\text{NIR}})\right|^2 \left|K_{\text{gap}}(\omega_{\text{MIR}})E_0(\omega_{\text{MIR}})\right|^2, \quad (S1)$$

where $\omega_{\text{MIR}}$ and $\omega_{\text{NIR}}$ represent the mid-IR and near-IR frequencies of excitation light for TE-SFG, respectively; and $\omega_{\text{SFG}}$ represents the sum frequency of $\omega_{\text{MIR}}$ and $\omega_{\text{NIR}}$. This equation provides a comprehensive expression for TE-SFG process that incorporates the spectral properties of the incident field enhancement ($K_{\text{gap}}$) and the signal emission efficiency ($L_{\text{gap}}$). In the main text, we denoted $\left|K_{\text{gap}}(\omega_{\text{MIR}})E_0(\omega_{\text{MIR}})\right|^2$ and $\left|K_{\text{gap}}(\omega_{\text{NIR}})E_0(\omega_{\text{NIR}})\right|^2$ as $I_{\text{MIR}}$ and $I_{\text{NIR}}$, respectively, and omitted $L_{\text{gap}}(\omega_{\text{SFG}})$ term for simplicity, arriving at the expression of eqs. (1) and (2) in the main text. Moreover, $\left|\chi^{(2)}\right|$ term in eq. (S1) represents the nonlinear optical response of sample materials placed within the tip–substrate nanogap. Therefore, to understand the fundamental mechanism governing the enhancement of SFG process within the nanogap, we particularly focus on $K_{\text{gap}}$ and $L_{\text{gap}}$ terms describing the field enhancement processes caused by the tip–substrate nanogap:

$$I_{\text{TESFG}}(\omega_{\text{SFG}}) \propto \left|L_{\text{gap}}(\omega_{\text{SFG}})\right|^2 \left|K_{\text{gap}}(\omega_{\text{NIR}})\right|^2 \left|K_{\text{gap}}(\omega_{\text{MIR}})\right|^2. \quad (S2)$$

Based on this equation, we next examine the spectral properties of the field enhancement factor ($K_{\text{gap}}$) and the enhanced emission efficiency ($L_{\text{gap}}$) to show the underlying enhancement mechanism of the TE-SFG process.

### 6.2 Computation of the field enhancement factor and radiation efficiency

The procedure for electromagnetic field simulations was described elsewhere.[1,25] Briefly, the finite-difference time-domain (FDTD) method was adopted with commercial software (Lumerical FDTD, Ansys). The system investigated in the simulation consists of a gold tip positioned above a gold substrate in vacuum, representing the nanogap in our STM (Figure 5a in the main text). The radius of tip apex, opening angle of the tip, and tip length were set to be 50 nm, 30°, and 15000 nm, respectively. These tip conditions correspond to the actual experimental conditions where the micrometer-scale Au tips (Figure 1c in the main text) were used. The refractive index of gold was taken from the experimental values of Olmon et al.[26] Perfectly matched layer boundary conditions were used in all simulations to absorb all outgoing waves and eliminate reflected light.

To evaluate the spectral properties of incident field enhancement ($K_{\text{gap}}$), we placed a monitor at the midpoint between the tip apex and the substrate surface to measure the electromagnetic field strength. A $p$-polarized Gaussian beam source with a waist of 2 μm was used to illuminate the nanogap at an incident of 55°. To evaluate the radiation efficiency ($L_{\text{gap}}$), an oscillating dipole source perpendicular to the gold substrate was placed at the same position as the monitor for $K_{\text{gap}}$. Radiated electromagnetic field from the dipole was monitored as the far-field at a position where the lateral and vertical distances from the dipole were 3000 and 2100 nm, respectively, matching the reflection angle of 55°.

### 6.3 Results
#### 6.3.1 Spectra of field enhancement factor and radiation efficiency

In this subsection, we summarize previously reported spectral characteristics of the field enhancement factor and radiation efficiency[1,25] to facilitate correct understandings of the results described in the main text. Figure S5a shows the calculated $|K_{\text{gap}}|^2$ spectrum. This spectrum is characterized by the broadband enhancement profile covering the visible and IR regions. As discussed in our previous work,[1,25] this broad enhancement is a clear manifestation of the influence of the micrometer-scale long tip shafts, whose origin can be attributed to the antenna effect caused by the collective oscillation of electrons over the entire tip. Notably, the magnified view of the $|K_{\text{gap}}|^2$ spectrum (Figure S5b) indicates that the value of $|K_{\text{gap}}|^2$ is almost constant across the experimental scan range of the mid-IR wavenumber (2800–3050 cm$^{-1}$).

On the other hand, the radiation process is not significantly influenced by the effect of tip shafts. As shown in Figure S5c, the spectra of $L_{\text{gap}}$ were still limited to a single band in the visible domain regardless of the tip length $l$. Although the tip length causes slight differences in the strength and shape of the $L_{\text{gap}}$ spectra, the wavelength range of efficient radiation from nanocavities is predominantly determined by gap-mode plasmons. This is because the time-averaged power of the vacuum propagating electromagnetic field radiated by the oscillating polarization is proportional to $\lambda^{-4}$ and steeply decreases with wavelength.

#### 6.3.2 Understanding the mechanisms of TE-SFG

The mechanisms of TE-SFG discussed below are also based on our recent studies,[1,25] which aid in interpreting the present results. The spectral characteristics of $K_{\text{gap}}$ and $L_{\text{gap}}$ are key to understand the mechanism of infrared-excited TE-SFG process. Figure S5d shows the wavelength dependence of the TE-SFG intensity calculated on the basis of eq. (S2) and $K_{\text{gap}}$ and $L_{\text{gap}}$ spectra shown in Figure S5a and c, respectively. In this calculation, $\omega_{\text{NIR}}$ was fixed to 9674.0 cm$^{-1}$ (1033.7 nm), which is the central frequency of the near-IR pulses used in our experiments (Figure S1a), and the TE-SFG intensity was plotted against $\omega_{\text{MIR}}$. The spectral profile indicates that TE-SFG process is highly efficient over broad IR range encompassing $\lambda \geq 1500$ nm region. This spectrally broad effectiveness of TE-SFG is the consequence of the fact that the broadband enhancement of incident light in the near-to-mid-IR region ($K_{\text{gap}}$) and efficient radiation in the visible-to-near-IR region ($L_{\text{gap}}$) effectively encompass the excitation and SFG radiation wavelength ranges, respectively. Thus, the SFG enhancement arises from the simultaneous amplification of excitation ($\omega_{\text{MIR}}$ and $\omega_{\text{NIR}}$) and radiation ($\omega_{\text{SFG}}$) processes caused by the concerted operation of two distinct enhancement mechanisms: the antenna effects caused by micrometer-scale tip shafts enhance infrared excitation, while localized gap-mode plasmons intensify the radiation at sum frequency.

Additionally, we expanded the tip–surface distance to 30 nm, which corresponds to the retracted condition in our experiments (the black curve in Figure 1f in the main text), and reexamined the spectra of $|K_{\text{gap}}|^2$ and $|L_{\text{gap}}|^2$ (Figure S6). As a result, the electric field enhancement ($K_{\text{gap}}$, Figure S6a) and emission efficiency ($L_{\text{gap}}$, Figure S6b) were found to be more than two orders of magnitude smaller than those in the tip-substrate plasmonic nanogap. These values should be too weak to produce a detectable enhancement of nonlinear optical signals from nanoscale tip apex, thereby allowing us to safely disregard the contribution from tip plasmon-enhanced signals in our experiments.

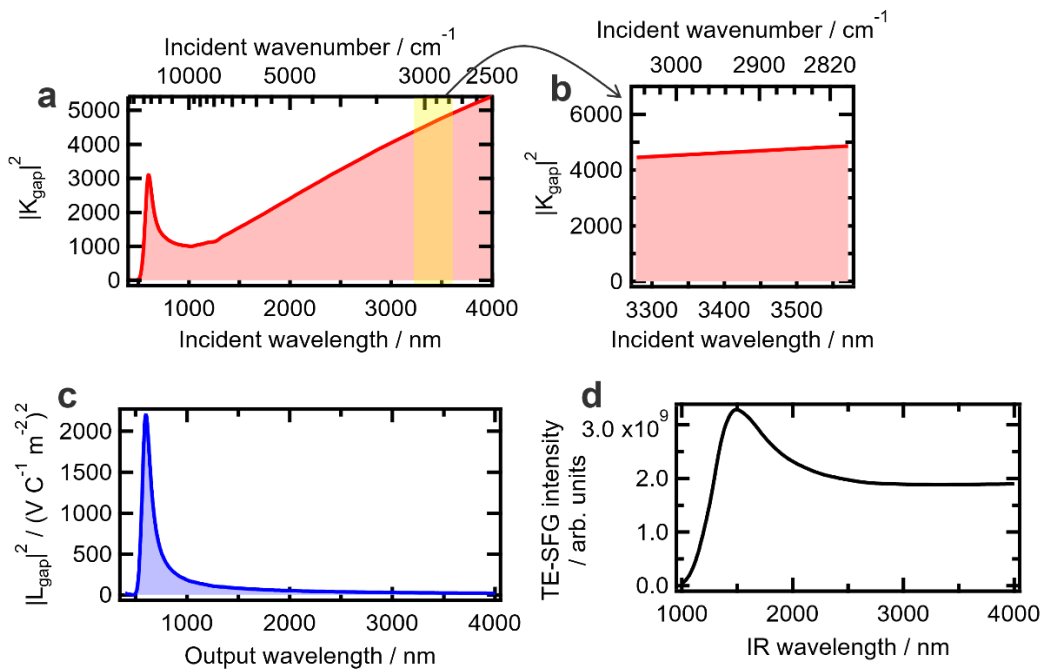

Figure S5. Theoretical calculation of the field enhancement factor and the emission efficiency. (a) Simulated $|K_{\text{gap}}|^2$ spectrum for a tip–substrate nanocavity obtained via FDTD calculations. (b) Magnified view of the yellow shaded region in panel (a). The wavelength (bottom axis) and corresponding wavenumber (top axis) ranges represent the scan range of the mid-IR excitation pulses used in our experiments. (c) Simulated $|L_{\text{gap}}|^2$ spectrum for a tip–substrate nanocavity obtained via FDTD calculations. (d) The excitation wavelength dependence of TE-SFG calculated based on eq. (S2) using $|K_{\text{gap}}|^2$ and $|L_{\text{gap}}|^2$ spectra shown in panels (a) and (c), respectively. In this calculation, one excitation wavelength was fixed at 1033 nm, corresponding to the central wavelength of near-IR excitation pulses used in our experiments. The horizontal axis represents the variable wavelength of another IR excitation light for SFG.

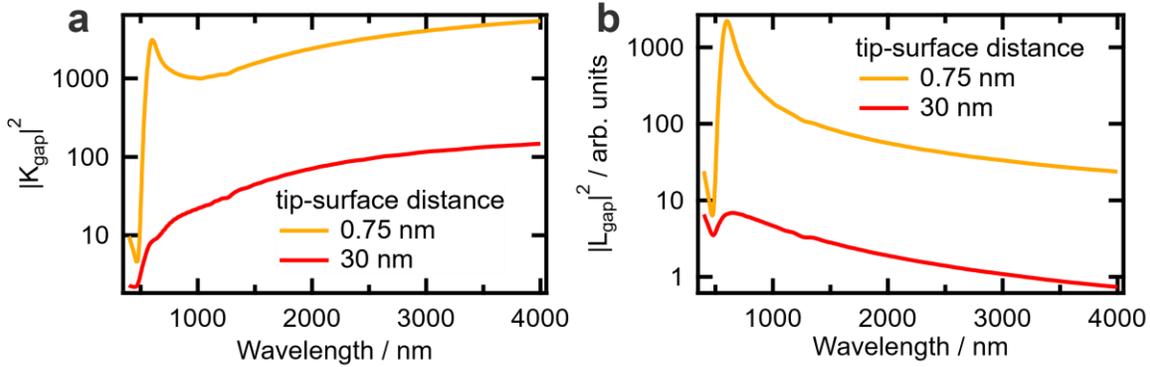

Figure S6. The spectra of field enhancement factor (a) and emission efficiency (b) calculated for tip–substrate gap distances of 0.75 nm (orange) and 30 nm (red).

## 7. Spatial distribution of electric fields

As discussed in the main text, the contribution of quadrupole transitions in the optical processes of molecules is determined by the relative ratio between the spatial scale of electric field amplitude variation and the size of an individual molecule. In this section, we quantitatively estimate the spatial distribution of the electric field within the nanogap through the FDTD method and evaluate the scale of electric field amplitude variation for comparison with molecular dimension.

Figure S7a and b display the surface-normal component of the electric field ($|E_Z|$), whereas Figure S7c and d display the intensity of surface-parallel component of the electric field ($\sqrt{E_X^2 + E_Y^2}$), which were calculated for a tip with a curvature radius of 50 nm The adopted tip radius is comparable to the size of the tip apex used in our experiments (Figure 1c in the main text). The surface-normal field is maximized in the vicinity of $(X, Y) = (0$ nm, $0$ nm$)$, where the tip–substrate gap distance is shortest (2 nm). Then, the field intensity gradually weakens as the distance from $X = 0$ nm position increases on the order of 10 nm. This behavior is more clearly visualized in Figure 5c in the main text, where the line profile of the field intensity along the $X$-axis with $Y = 0$ nm and $Z = 0.6$ nm is plotted. On the other hand, the surface-parallel component gets its maximum value around the side of the tip apex and is almost zero right underneath the tip apex. In addition, this surface-parallel field component is mainly distributed close to the tip surface and is minor in the vicinity of the substrate. As a result, the amplitude of the surface-parallel field along the $X$-axis with $Y = 0$ nm and $Z = 0.6$ nm (the red curve in Figure S7e) becomes significantly smaller than that of the surface-normal field (the green curve in Figure S7e). Therefore, surface-normal field dominantly contributes to the generation of $\boldsymbol{P}^{(2)}_{\mu\mu}$ within the tip–substrate nanogap (eq. (8) in the main text).

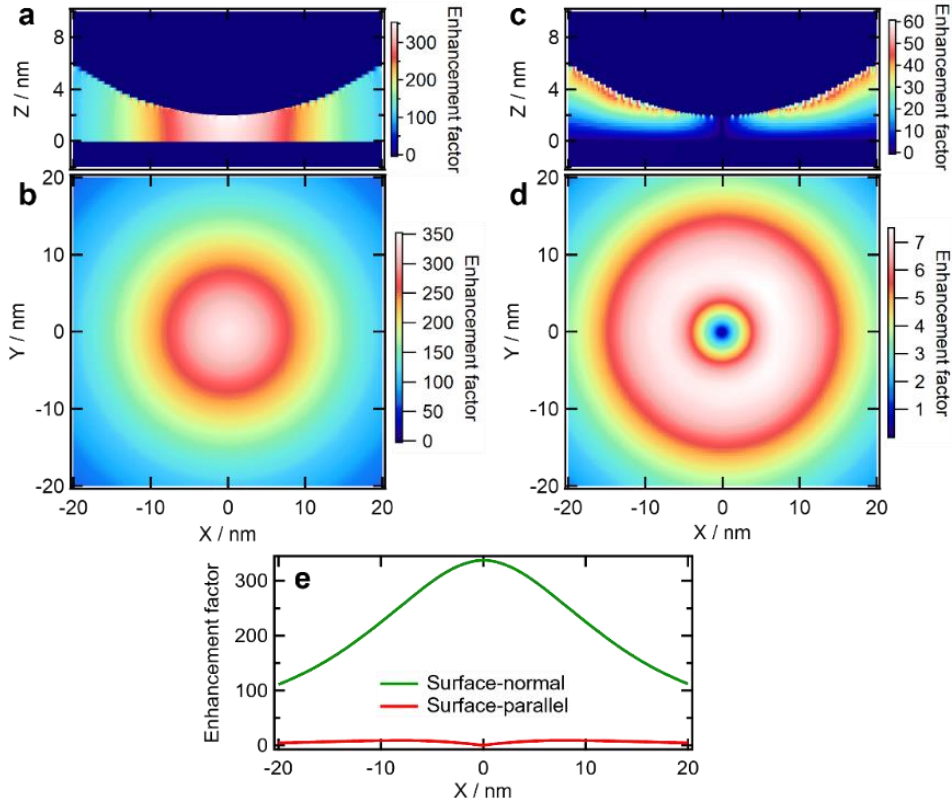

Figure S7. Spatial distributions of the field enhancement factor ($K_{gap}$) calculated through the FDTD simulations for (a, b) surface-normal and (c, d) surface-parallel electric field components. The model employs a rounded gold cone with a 30° opening angle, 50-nm apex radius, and 15000-nm length. The tip–substrate distance was set to be 2 nm. Panels (a) and (c) show the field in the $XZ$ plane ($Y$ = 0 nm), while panels (b) and (d) show the field in the $XY$ plane ($Z$ = 0.5 nm). The wavelength of incident light was 3425 nm. The coordinates ($X$, $Y$) = (0 nm, 0 nm) represent the position of minimum tip–substrate distance. Note that panel (a) is identical to Figure 5b in the main text. (e) One dimensional spatial profiles of $K_{gap}$ along the $X$-axis ($Y$ = 0 nm and $Z$ = 0.6 nm). Green and red curves represent the surface-normal and surface-parallel field components, respectively. The green curve is identical to that in Figure 5c in the main text.

## 8. Mechanism of SFG involving molecular dipole and quadrupole moments and the influence of electric field gradient

In this section, we overview the basic concepts of SFG process incorporating not only molecular dipole but also molecular quadrupole moments (Figure 4 in the main text). As briefly discussed in the main text, when the spatial variation of the electric field is negligible relative to molecular dimensions, the light–molecule interaction is primarily governed by the interaction between the molecular dipole moment and the electric field. In contrast, under specific conditions with a large field gradient comparable to molecular dimensions, additional contributions from the interaction between the molecular quadrupole moment and the electric field gradient[27–31] must be taken into account. This section provides the theoretical foundation for incorporating quadrupolar effects, along with the corresponding mathematical formulations of the SFG polarization that include both dipole- and quadrupole-induced contributions.

## 8.1 Induced dipole

In conventional far-field SFG measurements, incident electric fields induce oscillating dipole moments within individual molecules, and the coherent interference of radiation from a large number of such molecular dipoles ($\geq 10^6$) leads to a net directional SFG output that satisfies the phase-matching condition. In contrast, TE-SFG measurements involve a significantly reduced number of contributing molecules ($\leq 10^3$), rendering the phase-matching condition invalid. Accordingly, the TE-SFG electric field can be approximated by the dipolar radiation from the vector sum of the second-order molecular dipoles located beneath the tip apex ($\boldsymbol{P}^{(2)} = \sum_n \boldsymbol{p}_n^{(2),\text{mol}}$):

$$\boldsymbol{E}_{\text{p,SFG}} = k^2 \frac{\exp(ikr)}{r} \left(\boldsymbol{n} \times \boldsymbol{P}^{(2)}\right) \times \boldsymbol{n}, \tag{S3}$$

where $k = \omega/c$, $r = |\boldsymbol{r}|$, $\boldsymbol{n} = \boldsymbol{r}/r$. To understand the radiation characteristics associated with $\boldsymbol{P}^{(2)}$, we begin by examining the expression for the second-order nonlinear polarization of a single molecule $\boldsymbol{p}^{(2),\text{mol}}$.

When an individual molecule is irradiated by light, the induced polarization is given by:

$$\boldsymbol{p}^{\text{mol}} = \langle \Psi | \hat{\boldsymbol{\mu}} | \Psi \rangle, \tag{S4}$$

where $\hat{\boldsymbol{\mu}}$ is the operator of the dipole moment. The molecular wavefunction $\Psi$ can be perturbatively expanded with respect to the order of the light–matter interaction $\hat{V}$:

$$\Psi = \psi^{(0)} + \psi^{(1)} + \psi^{(2)} + \cdots. \tag{S5}$$

Substituting eq. (S5) into eq. (S4) yields multiple polarization terms corresponding to different orders of the perturbation. Among these terms, the second-order polarization relevant to SFG is expressed as:

$$\boldsymbol{p}^{(2),\text{mol}} = \langle \psi^{(0)} | \hat{\boldsymbol{\mu}} | \psi^{(2)} \rangle. \tag{S6}$$

This expression represents the molecular origin of the dipolar emission observed in SFG experiments. According to second-order perturbation theory, the relevant wavefunctions are given by

$$|\psi^{(0)}\rangle = |g\rangle, \qquad |\psi^{(2)}\rangle = \sum_e \sum_v \frac{\langle e|\hat{V}|v\rangle\langle v|\hat{V}|g\rangle}{(\omega_e - \omega_g)(\omega_v - \omega_g)} |e\rangle, \tag{S7}$$

where $|g\rangle$ is electronic and vibrational ground state of the molecule; and $\omega_g$, $\omega_v$, and $\omega_e$ are eigenfrequencies of states $|g\rangle$, $|v\rangle$, and $|e\rangle$, respectively. By substituting eq. (S7) into eq. (S6) and omitting the energy denominators for simplicity, the second-order molecular polarization simplifies to:

$$\boldsymbol{p}^{(2),\text{mol}} \propto \sum_e \sum_v \langle g|\hat{\boldsymbol{\mu}}|e\rangle\langle e|\hat{V}|v\rangle\langle v|\hat{V}|g\rangle. \tag{S8}$$

The interaction ($\hat{V}$) between an individual molecule and the incident electromagnetic field is expressed as the integral over all space of the product of the molecular charge density $\rho(\boldsymbol{r})$ and the electric potential $\phi(\boldsymbol{r})$ created by the incident field at position $\boldsymbol{r}$ in the molecular coordinate system:

$$\begin{aligned}
\hat{V} &= \int \rho(\boldsymbol{r})\phi(\boldsymbol{r})\,dV \\
&= \int \rho(\boldsymbol{r})\left(\phi(\boldsymbol{0}) + \sum_i^{x,y,z}\left(\frac{\partial\phi}{\partial i}\right)_0 i + \frac{1}{2}\sum_{i,j}^{x,y,z}\left(\frac{\partial^2\phi}{\partial i\partial j}\right)_0 ij + \cdots\right)dV \\
&\approx \left(\int \rho(\boldsymbol{r})\,dV\right)\phi(\boldsymbol{0}) + \sum_i^{x,y,z}\left(\int \rho(\boldsymbol{r})i\,dV\right)\left(\frac{\partial\phi}{\partial i}\right)_0 + \sum_{i,j}^{x,y,z}\frac{1}{2}\left(\int \rho(\boldsymbol{r})ij\,dV\right)\left(\frac{\partial^2\phi}{\partial i\partial j}\right)_0 \quad (S9) \\
&= Q\phi(\boldsymbol{0}) - \sum_i^{x,y,z}\mu_i\varepsilon_i(\boldsymbol{0}) - \sum_{i,j}^{x,y,z} q_{ij}\left(\frac{\partial\varepsilon_j}{\partial i}\right)_0 \\
&= Q\phi(\boldsymbol{0}) - \hat{\boldsymbol{\mu}}\cdot\boldsymbol{\varepsilon}(\boldsymbol{0}) - \hat{\boldsymbol{q}}{:}\nabla\boldsymbol{\varepsilon}(\boldsymbol{0}),
\end{aligned}$$

where $i, j, k, l$ denote the molecule-fixed coordinates $(x, y, z)$. $Q(=\int \rho(\boldsymbol{r})\,dV)$ represents the total charge, $\mu_i(=\int \rho(\boldsymbol{r})i\,dV)$ is the $i$ component of the dipole moment, and $q_{ij}(=(1/2)\int \rho(\boldsymbol{r})ij\,dV)$ is the $ij$ component of the quadrupole moment. In transitioning from the third to the fourth line, the relationship $\partial\phi/\partial i = -\varepsilon_i$ was employed. The operator ":" in the third term of the fifth line denotes the double inner product between two tensors, defined as $\boldsymbol{A}{:}\boldsymbol{B} = \sum_{i,j}^{x,y,z} A_{ij}B_{ij}$. The first term in the fifth line of eq. (S9) represents the electrostatic potential energy experienced by the charges within the molecule. For the subsequent discussion, we consider neutral molecules, allowing us to neglect this term. When the field gradient is sufficiently small, it is adequate to consider only the dipole–field interaction term $-\hat{\boldsymbol{\mu}}\cdot\boldsymbol{\varepsilon}(\boldsymbol{0})$. However, when strong field gradients exist on spatial scales comparable to molecular dimensions, it becomes necessary to include the quadrupole–field-gradient interaction term $-\hat{\boldsymbol{q}}{:}\nabla\boldsymbol{\varepsilon}(\boldsymbol{0})$. Substituting the relation $\hat{V} = -\hat{\boldsymbol{\mu}}\cdot\boldsymbol{\varepsilon}(\boldsymbol{0}) - \hat{\boldsymbol{q}}{:}\nabla\boldsymbol{\varepsilon}(\boldsymbol{0})$ into eq. (S8) yield the following expression:

$$\begin{aligned}
\boldsymbol{p}^{(2),\mathrm{mol}} &\propto \sum_e\sum_v \langle g|\hat{\boldsymbol{\mu}}|e\rangle\begin{bmatrix}(\langle e|\hat{\boldsymbol{\mu}}|v\rangle\cdot\boldsymbol{\varepsilon}(\boldsymbol{0}))(\langle v|\hat{\boldsymbol{\mu}}|g\rangle\cdot\boldsymbol{\varepsilon}(\boldsymbol{0})) \\ +(\langle e|\hat{\boldsymbol{q}}|v\rangle{:}\nabla\boldsymbol{\varepsilon}(\boldsymbol{0}))(\langle v|\hat{\boldsymbol{\mu}}|g\rangle\cdot\boldsymbol{\varepsilon}(\boldsymbol{0})) \\ +(\langle e|\hat{\boldsymbol{\mu}}|v\rangle\cdot\boldsymbol{\varepsilon}(\boldsymbol{0}))(\langle v|\hat{\boldsymbol{q}}|g\rangle{:}\nabla\boldsymbol{\varepsilon}(\boldsymbol{0})) \\ +(\langle e|\hat{\boldsymbol{q}}|v\rangle{:}\nabla\boldsymbol{\varepsilon}(\boldsymbol{0}))(\langle v|\hat{\boldsymbol{q}}|g\rangle{:}\nabla\boldsymbol{\varepsilon}(\boldsymbol{0}))\end{bmatrix} \\
&= \sum_e\sum_v\begin{bmatrix}\langle g|\hat{\boldsymbol{\mu}}|e\rangle(\langle e|\hat{\boldsymbol{\mu}}|v\rangle\cdot\boldsymbol{\varepsilon}(\boldsymbol{0}))(\langle v|\hat{\boldsymbol{\mu}}|g\rangle\cdot\boldsymbol{\varepsilon}(\boldsymbol{0})) \\ +\langle g|\hat{\boldsymbol{\mu}}|e\rangle(\langle e|\hat{\boldsymbol{q}}|v\rangle{:}\nabla\boldsymbol{\varepsilon}(\boldsymbol{0}))(\langle v|\hat{\boldsymbol{\mu}}|g\rangle\cdot\boldsymbol{\varepsilon}(\boldsymbol{0})) \\ +\langle g|\hat{\boldsymbol{\mu}}|e\rangle(\langle e|\hat{\boldsymbol{\mu}}|v\rangle\cdot\boldsymbol{\varepsilon}(\boldsymbol{0}))(\langle v|\hat{\boldsymbol{q}}|g\rangle{:}\nabla\boldsymbol{\varepsilon}(\boldsymbol{0})) \\ +\langle g|\hat{\boldsymbol{\mu}}|e\rangle(\langle e|\hat{\boldsymbol{q}}|v\rangle{:}\nabla\boldsymbol{\varepsilon}(\boldsymbol{0}))(\langle v|\hat{\boldsymbol{q}}|g\rangle{:}\nabla\boldsymbol{\varepsilon}(\boldsymbol{0}))\end{bmatrix}.
\end{aligned} \quad (S10)$$

In the following, we consider the $i$ component of $\boldsymbol{p}^{(2),\mathrm{mol}}$:

$$p_i^{(2),\text{mol}} \propto \sum_e \sum_v \begin{bmatrix} \sum_{j,k}^{x,y,z} \langle g|\hat{\mu}_i|e\rangle(\langle e|\hat{\mu}_j|v\rangle \varepsilon_j)(\langle v|\hat{\mu}_k|g\rangle \varepsilon_k) \\ + \sum_{j,k,l}^{x,y,z} \langle g|\hat{\mu}_i|e\rangle \left(\langle e|\hat{q}_{jl}|v\rangle \frac{\partial \varepsilon_j}{\partial l}\right)(\langle v|\hat{\mu}_k|g\rangle \varepsilon_k) \\ + \sum_{j,k,l}^{x,y,z} \langle g|\hat{\mu}_i|e\rangle(\langle e|\hat{\mu}_j|v\rangle \varepsilon_j)\left(\langle v|\hat{q}_{kl}|g\rangle \frac{\partial \varepsilon_k}{\partial l}\right) \\ + \sum_{j,k,l,m}^{x,y,z} \langle g|\hat{\mu}_i|e\rangle \left(\langle e|\hat{q}_{jl}|v\rangle \frac{\partial \varepsilon_j}{\partial l}\right)\left(\langle v|\hat{q}_{km}|g\rangle \frac{\partial \varepsilon_k}{\partial m}\right) \end{bmatrix}. \quad \text{(S11)}$$

Under the condition where a specific vibrational state resonates with an incident infrared light $p_i^{(2),\text{mol}}$ is given by

$$p_i^{(2),\text{mol}} \propto \sum_e \begin{bmatrix} \sum_{j,k}^{x,y,z} \langle g|\hat{\mu}_i|e\rangle(\langle e|\hat{\mu}_j|v\rangle \varepsilon_j)(\langle v|\hat{\mu}_k|g\rangle \varepsilon_k) \\ + \sum_{j,k,l}^{x,y,z} \langle g|\hat{\mu}_i|e\rangle \left(\langle e|\hat{q}_{jl}|v\rangle \frac{\partial \varepsilon_j}{\partial l}\right)(\langle v|\hat{\mu}_k|g\rangle \varepsilon_k) \\ + \sum_{j,k,l}^{x,y,z} \langle g|\hat{\mu}_i|e\rangle(\langle e|\hat{\mu}_j|v\rangle \varepsilon_j)\left(\langle v|\hat{q}_{kl}|g\rangle \frac{\partial \varepsilon_k}{\partial l}\right) \\ + \sum_{j,k,l}^{x,y,z} \langle g|\hat{\mu}_i|e\rangle \left(\langle e|\hat{q}_{jl}|v\rangle \frac{\partial \varepsilon_j}{\partial l}\right)\left(\langle v|\hat{q}_{km}|g\rangle \frac{\partial \varepsilon_k}{\partial m}\right) \end{bmatrix} \quad \text{(S12)}$$

$$= \sum_{j,k}^{x,y,z} \beta_{ijk}^{D\mu\mu,\text{mol}} \varepsilon_j \varepsilon_k + \sum_{j,k,l}^{x,y,z} \beta_{ijkl}^{Dq\mu,\text{mol}} \frac{\partial \varepsilon_j}{\partial l} \varepsilon_k + \sum_{j,k,l}^{x,y,z} \beta_{ijkl}^{D\mu q,\text{mol}} \varepsilon_j \frac{\partial \varepsilon_k}{\partial l}$$

$$+ \sum_{j,k,l,m}^{x,y,z} \beta_{ijklm}^{Dqq,\text{mol}} \frac{\partial \varepsilon_j}{\partial l} \frac{\partial \varepsilon_k}{\partial m}$$

$$= p_{i,\mu\mu}^{(2),\text{mol}} + p_{i,q\mu}^{(2),\text{mol}} + p_{i,\mu q}^{(2),\text{mol}} + p_{i,qq}^{(2),\text{mol}}.$$

The first term ($p_{i,\mu\mu}^{(2),\text{mol}}$) corresponds to polarization induced by "the dipole–field interaction of mid- and near-IR light". The second term ($p_{i,q\mu}^{(2),\text{mol}}$) represents polarization resulting from "the dipole–field interaction of mid-IR light" and "the quadrupole–field-gradient interaction of near-IR light". The third term ($p_{i,\mu q}^{(2),\text{mol}}$) denotes polarization arising from "the quadrupole–field-gradient interaction of mid-IR light" and "the dipole–field interaction of near-IR light". The fourth term ($p_{i,qq}^{(2),\text{mol}}$) is polarization induced by "the quadrupole–field-gradient interaction of mid- and near-IR light". Note that $\boldsymbol{P}_{\mu\mu}^{(2)}$, $\boldsymbol{P}_{q\mu}^{(2)}$, $\boldsymbol{P}_{\mu q}^{(2)}$, and $\boldsymbol{P}_{qq}^{(2)}$ introduced in Figure 4 in the main text are linked to

$p_{i,\mu\mu}^{(2),\text{mol}}$, $p_{i,q\mu}^{(2),\text{mol}}$, $p_{i,\mu q}^{(2),\text{mol}}$, and $p_{i,qq}^{(2),\text{mol}}$, respectively, by the relation $\boldsymbol{P}^{(2)} = \sum_n \boldsymbol{p}_n^{(2),\text{mol}}$. The hyperpolarizabilities for each term in the second line of eq. (S12) are expressed as follows:

$$\beta_{ijk}^{D\mu\mu,\text{mol}} = -\frac{1}{2\omega_a} \frac{\partial \alpha_{ij}}{\partial u_a} \frac{\partial \mu_k}{\partial u_a} \frac{1}{\omega_a - \omega_{\text{MIR}} - i\Gamma_a}, \tag{S13}$$

$$\beta_{ijkl}^{Dq\mu,\text{mol}} = -\frac{1}{2\omega_a} \frac{\partial \alpha'_{ijl}}{\partial u_a} \frac{\partial \mu_k}{\partial u_a} \frac{1}{\omega_a - \omega_{\text{MIR}} - i\Gamma_a}, \tag{S14}$$

$$\beta_{ijkl}^{D\mu q,\text{mol}} = -\frac{1}{2\omega_a} \frac{\partial \alpha_{ij}}{\partial u_a} \frac{\partial Q_{kl}}{\partial u_a} \frac{1}{\omega_a - \omega_{\text{MIR}} - i\Gamma_a}, \tag{S15}$$

$$\beta_{ijklm}^{Dqq,\text{mol}} = -\frac{1}{2\omega_a} \frac{\partial \alpha'_{ijl}}{\partial u_a} \frac{\partial Q_{km}}{\partial u_a} \frac{1}{\omega_a - \omega_{\text{MIR}} - i\Gamma_a}, \tag{S16}$$

where $\omega_a$ and $\Gamma_a$ are the angular frequency and the band width for the normal mode $a$, and $u_a$ is the vibrational coordinate for the mode $a$ with unit reduced mass. $\mu$, $Q$, $\alpha$, and $\alpha'$ are the dipole moment ($\langle v|\hat{\mu}|g\rangle$), quadrupole moment ($\langle v|\hat{q}|g\rangle$), dipole polarizability ($\sum_e \langle g|\hat{\mu}|e\rangle\langle e|\hat{\mu}|v\rangle$), and quadrupole polarizability ($\sum_e \langle g|\hat{\mu}|e\rangle\langle e|\hat{q}|v\rangle$), respectively. It is worth noting that the quadrupole-related terms incorporate not only electric quadrupole but also magnetic dipole.[31] Here, we introduced vibrational relaxation effects by including a phenomenological damping factor ($-i\Gamma_a$) in the energy denominators, which cannot be accounted for within the standard perturbative expansion of the wavefunction in eq. (S5).

To investigate the sensitivity of SFG signals to the absolute molecular orientation at interfaces, we focus on $p_{i,\mu\mu}^{(2),\text{mol}}$ and perform a coordinate transform from the molecular frame to the laboratory frame:

$$\begin{aligned} p_{I,\mu\mu}^{(2),\text{space}} &= \sum_i^{x,y,z} (\hat{I} \cdot \hat{i}) p_{i,\mu\mu}^{(2),\text{mol}} \\ &= \sum_i^{x,y,z} (\hat{I} \cdot \hat{i}) \sum_{j,k}^{x,y,z} \beta_{ijk}^{D\mu\mu,\text{mol}} \varepsilon_j \varepsilon_k \\ &= \sum_i^{x,y,z} (\hat{I} \cdot \hat{i}) \sum_{j,k}^{x,y,z} \beta_{ijk}^{D\mu\mu,\text{mol}} \left( \sum_J^{X,Y,Z} (\hat{J} \cdot \hat{j}) K_{\text{gap},J} E_J \right) \left( \sum_K^{X,Y,Z} (\hat{K} \cdot \hat{k}) K_{\text{gap},K} E_K \right) \\ &= \sum_{J,K}^{X,Y,Z} \left( \sum_{i,j,k}^{x,y,z} (\hat{I} \cdot \hat{i})(\hat{J} \cdot \hat{j})(\hat{K} \cdot \hat{k}) \beta_{ijk}^{D\mu\mu,\text{mol}} \right) K_{\text{gap},J} E_J K_{\text{gap},K} E_K, \end{aligned} \tag{S17}$$

where $(\hat{I} \cdot \hat{i})$, $(\hat{J} \cdot \hat{j})$, and $(\hat{K} \cdot \hat{k})$ represent the directional cosine components between molecular coordinate system $(i, j, k)$ and laboratory coordinate system $(I, J, K)$; and $K_{\text{gap},J}$ is $J$ component of the field enhancement factor introduced in section 6. Furthermore, the macroscopic dipolar polarization of the entire system can be represented by the sum of the dipolar polarization of individual molecules:

$$P_{I,\mu\mu}^{(2),\text{space}} = \sum_{n}^{N_{\text{mol}}} p_{n,I,\mu\mu}^{(2),\text{space}}$$

$$= \sum_{J,K}^{X,Y,Z} \left( N_{\text{mol}} \sum_{i,j,k}^{x,y,z} \langle (\hat{I} \cdot \hat{\imath})(\hat{J} \cdot \hat{\jmath})(\hat{K} \cdot \hat{k}) \rangle \beta_{ijk}^{D\mu\mu,\text{mol}} \right) K_{\text{gap},J} E_J K_{\text{gap},K} E_K. \quad (S18)$$

Here, $\langle (\hat{I} \cdot \hat{\imath})(\hat{J} \cdot \hat{\jmath})(\hat{K} \cdot \hat{k}) \rangle$ is the statistical average of directional cosines representing molecular orientations relative to laboratory coordinate system, and $N_{\text{mol}}$ is the number of molecules contributing to the signal. The terms in parenthesis in the second line serve as the coefficients linking the externally applied electric fields with the $I$ component of the macroscopic polarization of the system. We define this quantity as the second-order nonlinear susceptibility $\chi_{IJK}^{D\mu\mu,\text{space}}$:

$$\chi_{IJK}^{D\mu\mu,\text{space}} = N_{\text{mol}} \sum_{i,j,k}^{x,y,z} \langle (\hat{I} \cdot \hat{\imath})(\hat{J} \cdot \hat{\jmath})(\hat{K} \cdot \hat{k}) \rangle \beta_{ijk}^{D\mu\mu,\text{mol}}. \quad (S19)$$

This yields the well-known expression for the second-order nonlinear polarization:

$$P_{I,\mu\mu}^{(2),\text{space}} = \sum_{J,K}^{X,Y,Z} \chi_{IJK}^{D\mu\mu,\text{space}} K_{\text{gap},J} E_J K_{\text{gap},K} E_K. \quad (S20)$$

This corresponds to the dipolar term associated with the mechanism (a) illustrated in Figure 4 and eq. (8) in the main text.

In a similar manner, by applying the coordinate transformation to the dipolar polarizations arising from quadrupolar interactions ($p_{i,q\mu}^{(2),\text{mol}}$, $p_{i,\mu q}^{(2),\text{mol}}$, and $p_{i,qq}^{(2),\text{mol}}$), the macroscopic polarizations for the entire system can be expressed as follows:

$$P_{I,q\mu}^{(2),\text{space}} = \sum_{J,K,L}^{X,Y,Z} \left( N_{\text{mol}} \sum_{i,j,k,l}^{x,y,z} \langle (\hat{I} \cdot \hat{\imath})(\hat{J} \cdot \hat{\jmath})(\hat{K} \cdot \hat{k})(\hat{L} \cdot \hat{l}) \rangle \beta_{ijkl}^{Dq\mu,\text{mol}} \right) \frac{\partial (K_{\text{gap},J} E_J)}{\partial L} K_{\text{gap},K} E_K, \quad (S21)$$

$$P_{I,\mu q}^{(2),\text{space}} = \sum_{J,K,L}^{X,Y,Z} \left( N_{\text{mol}} \sum_{i,j,k,l}^{x,y,z} \langle (\hat{I} \cdot \hat{\imath})(\hat{J} \cdot \hat{\jmath})(\hat{K} \cdot \hat{k})(\hat{L} \cdot \hat{l}) \rangle \beta_{ijkl}^{D\mu q,\text{mol}} \right) K_{\text{gap},J} E_J \frac{\partial (K_{\text{gap},K} E_K)}{\partial L}, \quad (S22)$$

$$P_{I,qq}^{(2),\text{space}} = \sum_{J,K,L}^{X,Y,Z} \left( N_{\text{mol}} \sum_{i,j,k,l,m}^{x,y,z} \langle (\hat{I} \cdot \hat{\imath})(\hat{J} \cdot \hat{\jmath})(\hat{K} \cdot \hat{k})(\hat{L} \cdot \hat{l})(\hat{M} \cdot \hat{m}) \rangle \beta_{ijklm}^{Dqq,\text{mol}} \right)$$
$$\times \frac{\partial (K_{\text{gap},J} E_J)}{\partial L} \frac{\partial (K_{\text{gap},K} E_K)}{\partial M}. \quad (S23)$$

We define the terms in parentheses in eqs (S21), (S22), and (S23) as $\chi_{IJKL}^{Dq\mu,\text{space}}$, $\chi_{IJKL}^{D\mu q,\text{space}}$, and $\chi_{IJKL}^{Dqq,\text{space}}$, respectively:

$$\chi_{IJKL}^{Dq\mu,\text{space}} = N_{\text{mol}} \sum_{i,j,k,l}^{x,y,z} \langle (\hat{I} \cdot \hat{\imath})(\hat{J} \cdot \hat{\jmath})(\hat{K} \cdot \hat{k})(\hat{L} \cdot \hat{l}) \rangle \beta_{ijkl}^{Dq\mu,\text{mol}}, \quad (S24)$$

$$\chi_{IJKL}^{D\mu q,\text{space}} = N_\text{mol} \sum_{i,j,k,l}^{x,y,z} \langle(\hat{I}\cdot\hat{i})(\hat{J}\cdot\hat{j})(\hat{K}\cdot\hat{k})(\hat{L}\cdot\hat{l})\rangle \beta_{ijkl}^{D\mu q,\text{mol}}. \tag{S25}$$

$$\chi_{IJKLM}^{Dqq,\text{space}} = N_\text{mol} \sum_{i,j,k,l,m}^{x,y,z} \langle(\hat{I}\cdot\hat{i})(\hat{J}\cdot\hat{j})(\hat{K}\cdot\hat{k})(\hat{L}\cdot\hat{l})(\hat{M}\cdot\hat{m})\rangle \beta_{ijklm}^{Dqq,\text{mol}}. \tag{S26}$$

Through these definitions, the second-order nonlinear polarizations arising from quadrupole–field-gradient interactions can be compactly expressed as follows:

$$P_{I,q\mu}^{(2),\text{space}} = \sum_{J,K,L}^{X,Y,Z} \chi_{IJKL}^{Dq\mu,\text{space}} \frac{\partial(K_{\text{gap},J}E_J)}{\partial L} K_{\text{gap},K}E_K, \tag{S27}$$

$$P_{I,\mu q}^{(2),\text{space}} = \sum_{J,K,L}^{X,Y,Z} \chi_{IJKL}^{D\mu q,\text{space}} K_{\text{gap},J}E_J \frac{\partial(K_{\text{gap},K}E_K)}{\partial L}. \tag{S28}$$

$$P_{I,qq}^{(2),\text{space}} = \sum_{J,K,L,M}^{X,Y,Z} \chi_{IJKLM}^{Dqq,\text{space}} \frac{\partial(K_{\text{gap},J}E_J)}{\partial L} \frac{\partial(K_{\text{gap},K}E_K)}{\partial M}. \tag{S29}$$

Eqs. (S27), (S28), and (S29) correspond to the dipolar terms with mechanisms (b), (c), and (d) in Figure 4 in the main text, respectively. The summation of eqs. (S27) and (S28) corresponds to eq 9 in the main text. The resulting $\boldsymbol{P}_{\mu\mu}^{(2),\text{space}}$, $\boldsymbol{P}_{q\mu}^{(2),\text{space}}$, $\boldsymbol{P}_{\mu q}^{(2),\text{space}}$, and $\boldsymbol{P}_{qq}^{(2),\text{space}}$ serve as macroscopic sources for the SFG radiation. Notably, eqs. (S19) and (S23) contain three and five directional cosine components, respectively indicating that the absolute molecular orientation (up/down) information relative to the surface is directly encoded in the positive/negative sign of corresponding nonlinear susceptibilities $\chi_{IJK}^{D\mu\mu,\text{space}}$ and $\chi_{IJK}^{Dqq,\text{space}}$. In contrast, eqs. (S21) and (S22) involve four directional cosine components, indicating that absolute orientation information is not reflected in the signs of $\chi_{IJKL}^{Dq\mu,\text{space}}$ and $\chi_{IJKL}^{D\mu q,\text{space}}$. Therefore, accurate extraction of absolute molecular orientation from the measured SFG spectra requires quantifying the relative magnitudes of the dipolar and quadrupolar polarization contributions from $\boldsymbol{P}_{\mu\mu}^{(2),\text{space}}$, $\boldsymbol{P}_{q\mu}^{(2),\text{space}}$, $\boldsymbol{P}_{\mu q}^{(2),\text{space}}$, and $\boldsymbol{P}_{qq}^{(2),\text{space}}$.

### 8.2 Induced quadrupole

In the previous subsection, we focused on molecular dipolar polarization that is optically induced through dipole–field and quadrupole–field-gradient interactions. Importantly, however, molecular polarizations caused by the light–molecule interaction are not limited to dipolar polarizations; higher-order polarizations can also be generated and contribute to field radiation. In this subsection, among such higher-order polarizations, we particularly focus on the contribution of quadrupole radiation and derive expressions for the nonlinear susceptibility governing the generation of quadrupole moments.

Analogous to the dipole case, the TE-SFG electric field emitted from quadrupolar polarizations can be approximated as

$$\boldsymbol{E}_{\text{Q,SFG}} = -\mathrm{i}k^3 \frac{\exp(\mathrm{i}kr)}{r}\left(\boldsymbol{n} \times \left(\boldsymbol{Q}^{(2)} \cdot \boldsymbol{n}\right)\right) \times \boldsymbol{n}, \tag{S30}$$

where $\boldsymbol{Q}^{(2)}$ represents the vector sum of molecular quadrupoles located beneath the tip apex $\left(\boldsymbol{Q}^{(2)} = \sum_n \boldsymbol{q}_n^{(2),\text{mol}}\right)$. To understand the radiation characteristics associated with $\boldsymbol{Q}^{(2)}$, we begin by examining the second-order quadrupolar polarization of a single molecule $\boldsymbol{q}^{(2)}$.

The optically induced second-order quadrupole moment of individual molecules is defined as

$$\boldsymbol{q}^{(2),\text{mol}} = \langle \psi^{(0)} | \hat{\boldsymbol{q}} | \psi^{(2)} \rangle, \tag{S31}$$

where $\hat{\boldsymbol{q}}$ is the operator of the quadrupole moment. Considering second-order interactions with light (double interactions with $\hat{V}$), the induced quadrupole moment can be expressed as

$$\boldsymbol{q}^{(2),\text{mol}} \propto \sum_e \sum_v \langle g|\hat{\boldsymbol{q}}|e\rangle \langle e|\hat{V}|v\rangle \langle v|\hat{V}|g\rangle. \tag{S32}$$

Similarly to the dipole case, this expression comprises four distinct contributions induced by different interaction pathways: (i) dipole interactions with mid- and near-IR electric fields, (ii) dipole interactions with the mid-IR electric field and quadrupole interactions with the near-IR field gradient, (iii) dipole interactions with the near-IR electric field and quadrupole interactions with the mid-IR field gradient, and (iv) quadrupole interactions with mid- and near-IR electric fields gradients:

$$\begin{aligned}
q_{ij}^{(2),\text{mol}} &= \sum_{k,l}^{x,y,z} \beta_{ijkl}^{Q\mu\mu,\text{mol}} \varepsilon_k \varepsilon_l + \sum_{k,l,m}^{x,y,z} \beta_{ijklm}^{Qq\mu,\text{mol}} \frac{\partial \varepsilon_k}{\partial m} \varepsilon_l + \sum_{k,l,m}^{x,y,z} \beta_{ijklm}^{Q\mu q,\text{mol}} \varepsilon_k \frac{\partial \varepsilon_l}{\partial m} \\
&\quad + \sum_{k,l,m,n}^{x,y,z} \beta_{ijklmn}^{Qqq,\text{mol}} \frac{\partial \varepsilon_k}{\partial m} \frac{\partial \varepsilon_l}{\partial n} \\
&= q_{ij,\mu\mu}^{(2),\text{mol}} + q_{ij,q\mu}^{(2),\text{mol}} + q_{ij,\mu q}^{(2),\text{mol}} + q_{ij,qq}^{(2),\text{mol}},
\end{aligned} \tag{S33}$$

where $\beta_{ijkl}^{Q\mu\mu,\text{mol}}$, $\beta_{ijklm}^{Qq\mu,\text{mol}}$, $\beta_{ijklm}^{Q\mu q,\text{mol}}$, and $\beta_{ijklmn}^{Qqq,\text{mol}}$ are the hyperpolarizabilities associated with the molecular quadrupole moments $q_{ij,\mu\mu}^{(2),\text{mol}}$, $q_{ij,q\mu}^{(2),\text{mol}}$, $q_{ij,\mu q}^{(2),\text{mol}}$, and $q_{ij,qq}^{(2),\text{mol}}$. Note that $\boldsymbol{Q}_{\mu\mu}^{(2)}$, $\boldsymbol{Q}_{q\mu}^{(2)}$, $\boldsymbol{Q}_{\mu q}^{(2)}$, and $\boldsymbol{Q}_{qq}^{(2)}$ introduced in Figure 4 in the main text correspond to $q_{ij,\mu\mu}^{(2),\text{mol}}$, $q_{ij,q\mu}^{(2),\text{mol}}$, $q_{ij,\mu q}^{(2),\text{mol}}$, and $q_{ij,qq}^{(2),\text{mol}}$, respectively. Transforming these quadrupolar terms into the laboratory coordinate system yields

$$q_{IJ,\mu\mu}^{(2),\text{space}} = \sum_{K,L}^{X,Y,Z} \left( \sum_{i,j,k,l}^{x,y,z} (\hat{I}\cdot\hat{i})(\hat{J}\cdot\hat{j})(\hat{K}\cdot\hat{k})(\hat{L}\cdot\hat{l}) \beta_{ijkl}^{Q\mu\mu,\text{mol}} \right) K_{\text{gap},K} E_K K_{\text{gap},L} E_L, \tag{S34}$$

$$\begin{aligned}
q_{IJ,q\mu}^{(2),\text{space}} &= \sum_{K,L,M}^{X,Y,Z} \left( \sum_{i,j,k,l,m}^{x,y,z} (\hat{I}\cdot\hat{i})(\hat{J}\cdot\hat{j})(\hat{K}\cdot\hat{k})(\hat{L}\cdot\hat{l})(\hat{M}\cdot\hat{m}) \beta_{ijklm}^{Qq\mu,\text{mol}} \right) \\
&\quad \times \frac{\partial (K_{\text{gap},K} E_K)}{\partial M} K_{\text{gap},L} E_L,
\end{aligned} \tag{S35}$$

$$q_{IJ,\mu q}^{(2),\text{space}} = \sum_{K,L,M}^{X,Y,Z} \left( \sum_{i,j,k,l,m}^{x,y,z} (\hat{I}\cdot\hat{i})(\hat{J}\cdot\hat{j})(\hat{K}\cdot\hat{k})(\hat{L}\cdot\hat{l})(\hat{M}\cdot\hat{m})\beta_{ijklm}^{Q\mu q,\text{mol}} \right)$$
$$\times K_{\text{gap},K} E_K \frac{\partial(K_{\text{gap},L}E_L)}{\partial M}, \tag{S36}$$

$$q_{IJ,\mu q}^{(2),\text{space}} = \sum_{K,L,M,N}^{X,Y,Z} \left( \sum_{i,j,k,l,m,n}^{x,y,z} (\hat{I}\cdot\hat{i})(\hat{J}\cdot\hat{j})(\hat{K}\cdot\hat{k})(\hat{L}\cdot\hat{l})(\hat{M}\cdot\hat{m})(\hat{M}\cdot\hat{m})\beta_{ijklmn}^{Qqq,\text{mol}} \right)$$
$$\times \frac{\partial(K_{\text{gap},K}E_K)}{\partial M}\frac{\partial(K_{\text{gap},L}E_L)}{\partial N}, \tag{S37}$$

The macroscopic quadrupole moment for the entire system can be represented as the sum of individual molecular quadrupole moments

$$Q_{IJ}^{(2),\text{space}} = \sum_{n}^{N} q_{n,IJ}^{(2),\text{space}} \tag{S38}$$

Thus, taking the statistical average of the terms in parentheses from eqs. (S34)–(S37) and multiplying by the number of molecules $N_{\text{mol}}$ yields the nonlinear susceptibilities describing the macroscopic quadrupole generation:

$$Q_{IJ,\mu\mu}^{(2),\text{space}} = N_{\text{mol}} \sum_{K,L}^{X,Y,Z} \chi_{IJKL}^{Q\mu\mu,\text{space}} K_{\text{gap},K}E_K K_{\text{gap},L}E_L \tag{S39}$$

$$Q_{IJ,q\mu}^{(2),\text{space}} = N_{\text{mol}} \sum_{K,L,M}^{X,Y,Z} \chi_{IJKLM}^{Qq\mu,\text{space}} \frac{\partial(K_{\text{gap},K}E_K)}{\partial M} K_{\text{gap},L}E_L \tag{S40}$$

$$Q_{IJ,\mu q}^{(2),\text{space}} = N_{\text{mol}} \sum_{K,L,M}^{X,Y,Z} \chi_{IJKLM}^{Q\mu q,\text{space}} K_{\text{gap},K}E_K \frac{\partial(K_{\text{gap},L}E_L)}{\partial M} \tag{S41}$$

$$Q_{IJ,qq}^{(2),\text{space}} = N_{\text{mol}} \sum_{K,L,M,N}^{X,Y,Z} \chi_{IJKLMN}^{Qqq,\text{space}} \frac{\partial(K_{\text{gap},K}E_K)}{\partial M}\frac{\partial(K_{\text{gap},L}E_L)}{\partial N} \tag{S42}$$

Eqs. (S39)–(S42) correspond to the quadrupolar terms induced through mechanisms (a)–(d) in Figure 4 in the main text, respectively. Similar to $\chi_{IJKL}^{Dq\mu,\text{space}}$ and $\chi_{IJKL}^{D\mu q,\text{space}}$, the susceptibilities $\chi_{IJKL}^{Q\mu\mu,\text{space}}$ and $\chi_{IJKLMN}^{Qqq,\text{space}}$ contain even-number directional cosine components, and thus are insensitive to the absolute molecular orientation. In contrast, $\chi_{IJKLM}^{Qq\mu,\text{space}}$ and $\chi_{IJKLM}^{Q\mu q,\text{space}}$ contain five directional cosine components, thereby involving the sensitivity to the absolute molecular orientation.

## 9. Evaluation of dipolar and quadrupolar contributions in the TE-SFG signals

In the previous section, we discussed the multiple mechanisms contributing to the TE-SFG processes including dipolar polarizations (eqs. (S20), (S27)–(S29)) and the quadrupolar polarizations (eqs. (S39)–(S42)). In this section, we quantitatively evaluate the relative contributions of these mechanisms to the overall TE-SFG intensity by combining the simulated spatial field distribution obtained from FDTD calculations (section 7) with quantum chemical calculations. Our analysis confirms that, under current experimental conditions, the contribution of the quadrupole effects is negligible in our TE-SFG signals.

### 9.1 Dipolar radiation
#### 9.1.1 Polarizations and hyperpolarizabilities

We first focus on dipolar polarizations induced through mechanisms (a)–(c) illustrated in Figure 4 in the main text ($\boldsymbol{P}^{(2),\text{space}}_{\mu\mu}$, $\boldsymbol{P}^{(2),\text{space}}_{q\mu}$, and $\boldsymbol{P}^{(2),\text{space}}_{\mu q}$) and the corresponding SFG radiation from them. The net dipolar polarizations formed from those three polarizations can be explicitly expressed by using eqs S20, S27, and S28:

$$
\begin{aligned}
P_I^{(2),\text{space}}(\Omega) &= P_{I,\mu\mu}^{(2),\text{space}} + P_{I,q\mu}^{(2),\text{space}} + P_{I,\mu q}^{(2),\text{space}} \\
&= \sum_{J,K}^{X,Y,Z} \chi_{IJK}^{D\mu\mu,\text{space}} K_{\text{gap},J} E_J K_{\text{gap},K} E_K \\
&\quad + \sum_{J,K,L}^{X,Y,Z} \chi_{IJKL}^{Dq\mu,\text{space}} \frac{\partial\left(K_{\text{gap},J} E_J\right)}{\partial L} K_{\text{gap},K} E_K \\
&\quad + \sum_{J,K,L}^{X,Y,Z} \chi_{IJKL}^{D\mu q,\text{space}} K_{\text{gap},J} E_J \frac{\partial\left(K_{\text{gap},K} E_K\right)}{\partial L},
\end{aligned}
\tag{S43}
$$

The spatial distribution of electric fields and their gradients contained in this equation were obtained from the FDTD calculations detailed in section 7. To quantitatively calculate TE-SFG intensities, the explicit values of the second-order nonlinear susceptibilities ($\chi_{IJKL}^{D\mu\mu,\text{space}}$, $\chi_{IJKL}^{Dq\mu,\text{space}}$, and $\chi_{IJKL}^{D\mu q,\text{space}}$) are required. As shown in eqs. (S19), (S24), and (S25), these susceptibilities are related to the molecular hyperpolarizabilities in the molecular-fixed coordinate system ($\beta_{ijk}^{D\mu\mu,\text{mol}}$, $\beta_{ijkl}^{Dq\mu,\text{mol}}$, and $\beta_{ijkl}^{D\mu q,\text{mol}}$) and orientational distribution of the molecules expressed by the statistical average of directional cosines between the molecular coordinate system and laboratory coordinate system (Figure S8). The hyperpolarizabilities $\beta_{ijk}^{D\mu\mu,\text{mol}}$, $\beta_{ijkl}^{Dq\mu,\text{mol}}$, and $\beta_{ijkl}^{D\mu q,\text{mol}}$ for each vibrational mode $a$ in the molecule-fixed coordinates are given by eqs. (S13), (S14), and (S15), respectively. These hyperpolarizabilities were calculated for a 4-MBT molecule by using quantum susceptibility automatic calculator (Qsac), which was developed in our previous study.[27] Figure S8 illustrates the molecule-fixed coordinates for 4-MBT, where the $z$ axis is along the C-C vector between the phenyl ring and the methyl group and the $y$ axis is normal to the benzene ring. The vibrational analysis has already been performed in Section 5 to obtain $\omega_a$ and $u_a$ for each mode $a$. The mode-specific derivatives $\partial G/\partial u_a$ ($G = \mu, Q, \alpha, \alpha'$) in eqs. (S13)–(S15) were calculated with the same level. The major derivative components for the methyl C-H symmetric stretching mode are shown in Table S3.

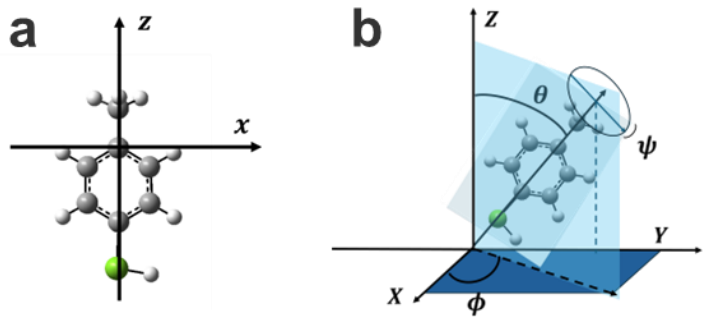

Figure S8 (a) Molecule-fixed coordinates of 4-MBT and (b) orientation of 4-MBT molecule with Euler angles ($\phi, \theta, \psi$) defined according to the $z - y - z$ rotation convention.

Table S3 The major components of $\partial G/\partial u_a$ ($G = \mu, Q, \alpha, \alpha'$) for the methyl C-H symmetric stretching mode $a$ of a 4-MBT molecule. The molecule-fixed coordinates ($x, y, z$) are defined in Figure S8a. Unit: atomic units.

| $G$ | (a.u.) | $G$ | (a.u.) | $G$ | (a.u.) |
|---|---|---|---|---|---|
| $\mu_z$ | $-0.0044$ | $Q_{xx}$ | $-0.0034$ | $\alpha'_{xzx}$ | $0.5447$ |
| $\alpha_{xx}$ | $0.0988$ | $Q_{yy}$ | $-0.0045$ | $\alpha'_{yyy}$ | $-0.1413$ |
| $\alpha_{yy}$ | $0.1697$ | $Q_{zz}$ | $-0.0083$ | $\alpha'_{yzy}$ | $0.8089$ |
| $\alpha_{zz}$ | $0.3539$ | $Q_{yz}$ | $0.0059$ | $\alpha'_{yyz}$ | $0.1688$ |
| | | | | $\alpha'_{zzz}$ | $0.2034$ |

### 9.1.2 Calculation of induced dipolar polarization

We calculated the dipole polarizations of a 4-MBT molecule placed at various lateral positions on the surface. The dipole polarizations are represented in eq. (S43), where the electric field $K_{\text{gap}}E$ and its gradient $\partial(K_{\text{gap}}E)/\partial L$ are functions of $XYZ$ position on the surface while $\chi_{IJKL}^{D\mu\mu,\text{space}}$, $\chi_{IJKL}^{Dq\mu,\text{space}}$, and $\chi_{IJKL}^{D\mu q,\text{space}}$ depend on the orientation with directional cosine terms in eqs. (S19), (S24), and (S25). The lateral positions were divided by grid with 1-nm interval, and the electric field $K_{\text{gap}}E$ and its gradient $\partial(K_{\text{gap}}E)/\partial L$ at each grid point were obtained from the field distribution calculated through FDTD method (section 7). The orientation is represented with the set of Euler angles ($\phi, \theta, \psi$) in Figure S8b. We examined three cases with tilt angles of $\theta = 0°, 30°, 60°$, while the other two angles $\phi, \psi$ were assumed to be uniformly distributed. The average of rotation matrix products $\langle(\hat{I}\cdot\hat{\imath})(\hat{J}\cdot\hat{\jmath})(\hat{K}\cdot\hat{k})\rangle$ in eq. (S19) and $\langle(\hat{I}\cdot\hat{\imath})(\hat{J}\cdot\hat{\jmath})(\hat{K}\cdot\hat{k})(\hat{L}\cdot\hat{l})\rangle$ in eqs. (S24) and (S25) were calculated in each case of orientation.

### 9.1.3 Intensity of dipolar SFG radiation

The radiation emitted from the dipole polarizations in eq. (S43) was evaluated. The radiation intensities from the dipole over the whole solid angle and band frequency are given by

$$I[P^{(2)}(\Omega)(E)] = \frac{c}{3}\left(\frac{n\Omega}{c}\right)^4 \int d\omega_{\text{MIR}} \sum_I |P_I(\Omega, E)|^2, \tag{S44}$$

$$I[P^{(2)}(\Omega)(\nabla E)] = \frac{c}{3}\left(\frac{n\Omega}{c}\right)^4 \int d\omega_{\text{MIR}} \sum_I |P_I(\Omega, \nabla E)|^2, \tag{S45}$$

where $c$ is the speed of light in vacuum; $n$ is the refractive index of the media; $P_I(\Omega, E)$ is the first term in eq. (S43) ($P_{I,\mu\mu}^{(2),\text{space}}$); $P_I(\Omega, \nabla E)$ is the sum of the second and third terms in eq. (S43) ($P_{I,q\mu}^{(2),\text{space}} + P_{I,\mu q}^{(2),\text{space}}$).

Figure S9 displays two-dimensional maps of calculated dipolar SFG intensity from 4-MBT at various lateral positions (X, Y) on Au substrate, where the tilt angle of 4-MBT is $\theta = 0°$. These pictures reveal two important features of TE-SFG spectroscopy. First, the spot of high intensity extends up to ~10 nm from the origin, implying that the TE-SFG in the present configuration has a lateral resolution in this spatial range. Indeed, the spatial resolution of TE-SFG experimentally demonstrated in the main text is in the same order as this estimation. Second, the TE-SFG signal is dominated by the radiation from the dipolar polarization induced by the electric field ($\boldsymbol{P}_{\mu\mu}^{(2),\text{space}}$), like the conventional far-field SFG. Comparing the SFG intensities from $\boldsymbol{P}_{\mu\mu}^{(2),\text{space}}$ (Figure S9a) and $\boldsymbol{P}_{q\mu}^{(2),\text{space}} + \boldsymbol{P}_{\mu q}^{(2),\text{space}}$ (Figure S9b), we find that $\boldsymbol{P}_{q\mu}^{(2),\text{space}} + \boldsymbol{P}_{\mu q}^{(2),\text{space}}$ exhibits significantly weaker intensity than $\boldsymbol{P}_{\mu\mu}^{(2),\text{space}}$ by a factor of $10^{-5}$. Moreover, given that replacing only one of the two interactions with a quadrupole–field-gradient interaction results in such a dramatic reduction, the contribution from double quadrupole interactions, $\boldsymbol{P}_{qq}^{(2),\text{space}}$, must be even smaller. Therefore, under the current tip conditions with ~50-nm radius of apex, $\boldsymbol{P}_{\mu\mu}^{(2),\text{space}}$ dominantly contributes to the observed signals, and the effects of the quadrupole–field-gradient interactions can be safely ruled out in the dipolar SFG radiation.

### 9.2 Quadrupolar radiation

We next focus on quadrupolar polarizations. As demonstrated in the previous subsection, the SFG processes involving quadrupole–field-gradient interactions negligibly contribute to the overall SFG signal radiation under the current tip conditions. Thus, we feature only $\boldsymbol{Q}_{\mu\mu}^{(2),\text{space}}$ term presented in eq. (S39) and quantitatively evaluate its contribution.

The $\chi_{IJKL}^{Q\mu\mu,\text{space}}$ term in eq. (S39) is given by

$$\chi_{IJKL}^{Q\mu\mu,\text{space}} = \sum_{i,j,k,l}^{x,y,z} \langle (\hat{I}\cdot\hat{i})(\hat{J}\cdot\hat{j})(\hat{K}\cdot\hat{k})(\hat{L}\cdot\hat{l})\rangle \beta_{ijkl}^{Q\mu\mu,\text{mol}}, \tag{S46}$$

where the hyperpolarizability $\beta_{ijkl}^{Q\mu\mu,\text{mol}}$ is expressed as

$$\beta_{ijkl}^{Q\mu\mu,\text{mol}} = -\frac{1}{2\omega_a}\frac{\partial \alpha'_{kij}}{\partial u_a}\frac{\partial \mu_l}{\partial u_a}\frac{1}{\omega_a - \omega_{\text{MIR}} - i\Gamma_a}. \tag{S47}$$

The explicit values of $\partial \alpha'_{kij}/\partial u_a$ and $\partial \mu_l/\partial u_a$ terms in this equation have already been calculated (Table S3). Moreover, the radiation intensities from the quadrupole over the whole solid angle and band frequency are given by

$$I[Q^{(2)}(\Omega)(E)] = \frac{c}{30}\left(\frac{n\Omega}{c}\right)^6 \int d\omega_{\text{MIR}} \left\{ \begin{array}{c} 5\sum_{I,J}|Q_{IJ}(\Omega)|^2 - \frac{1}{2}\sum_{I,J}|Q_{IJ}(\Omega) + Q_{JI}(\Omega)|^2 \\ + \left|\sum_I Q_{II}(\Omega)\right|^2 \end{array} \right\}. \tag{S48}$$

By using these relationships, we calculated the spatial distribution of SFG intensity originating from quadrupolar polarization in a manner analogous to the dipole case. As shown in Figure S9c, the contribution from $\boldsymbol{Q}_{\mu\mu}^{(2),\text{space}}$ is seven orders of magnitude smaller than that from $\boldsymbol{P}_{\mu\mu}^{(2),\text{space}}$, clearly indicating that the quadrupole emission is negligibly small. A comparison of Figure S9a–c confirms that the effects of the molecular quadrupole moments and the electric field gradients can be safely ruled out, thereby validating the applicability of the electric dipole approximation under our experimental conditions. Note that the simulated SFG intensities presented in Figure 5e in the main text correspond to line profiles along the $X$-axis at $Y = 0$ nm. Importantly, the dominance of the dipole radiation mechanism remains valid for other molecular tilt angles of 4-MBT (Figure S10).

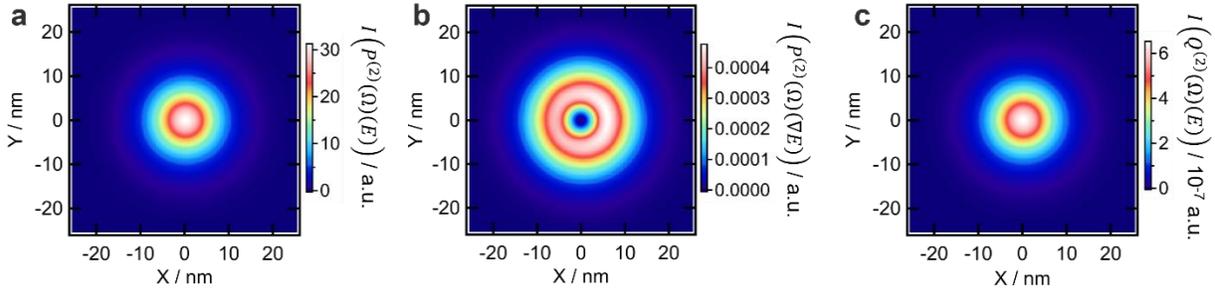

Figure S9 Two-dimensional maps of TE-SFG intensity emitted from (a) $\boldsymbol{P}_{\mu\mu}^{(2),\text{space}}$, (b) $\boldsymbol{P}_{q\mu}^{(2),\text{space}} + \boldsymbol{P}_{\mu q}^{(2),\text{space}}$, and $\boldsymbol{Q}_{\mu\mu}^{(2),\text{space}}$ originating from 4-MBT molecules on Au substrate. The two-dimensional distributions in panels (a), (b), and (c) were calculated according to eqs. (S44), (S45), and (S48), respectively. The intensities are plotted as a function of lateral position ($X$, $Y$), where (0,0) corresponds to the position directly beneath the tip center. The orientation of methyl group is assumed to be $\theta = 0°$. Although the values are plotted in arbitrary units, the intensities in (a–c) are directly comparable.

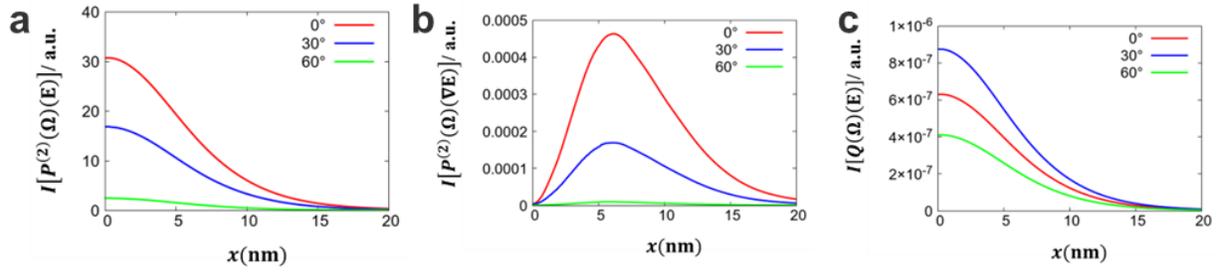

Figure S10 SFG intensities from 4-MBT at ($X$, $Y$=0) as a function of $X$. Three cases of tilt angles $\theta = 0°$, $30°$, $60°$ are plotted in each panel of three mechanisms (a–c). The unit of the ordinate in each panel is common with the contour of the corresponding panel in Figure S9.

## 10. Relation between molecular orientation and $\mathrm{Im}\left(\chi_\mathrm{R}^{(2)}\right)$ signals

### 10.1 General expression of $\chi_\mathrm{R}^{(2)}$

In this section, we show that the sign of $\mathrm{Im}\left(\chi_\mathrm{R}^{(2)}\right)$ signals obtained through TE-SFG experiments (Figure 3c and d in the main text) is directly related to the absolute up/down orientation of 4-MBT molecules. As discussed in sections 7–9, the electric field gradient at the tip–substrate nanogap realized in the present study is negligibly small compared with the size of individual molecules. Thus, the following discussion on the relation between molecular orientation and $\mathrm{Im}\left(\chi_\mathrm{R}^{(2)}\right)$ signals is based on a well-established far-field SFG theory under the electric dipole approximation for the optical responses of molecules.

We redefine $\chi_{IJK}^{D\mu\mu,\mathrm{space}}$ in eq. (S19) as $\chi_{IJK}^{(2)}$ for simplicity. Among the total of 27 tensor elements of $\chi_{IJK}^{(2)}$, there are seven non-zero terms for an achiral rotationally isotropic interface, namely, $\chi_{XXZ}^{(2)} = \chi_{YYZ}^{(2)}$, $\chi_{XZX}^{(2)} = \chi_{YZY}^{(2)}$, $\chi_{ZXX}^{(2)} = \chi_{ZYY}^{(2)}$, $\chi_{ZZZ}^{(2)}$.[32] This assumption of configurational symmetry is valid under our experimental conditions, where the molecules are uniformly adsorbed on the surface and the conical tip is placed above the surface with its axis parallel to the surface normal. In TE-SFG process, out of the seven non-zero tensor elements, only $\chi_{ZZZ}^{(2)}$ play dominant roles because intense electric field enhancement works effectively only for $Z$-directed field and the field components parallel to the metal surface are significantly smaller than the $Z$-directed field component (see Section 7 for further details). Hereafter, we denote this $\chi_{ZZZ}^{(2)}$ as just $\chi_\mathrm{R}^{(2)}$ for simplicity.

The vibrationally resonant term of $\chi_\mathrm{R}^{(2)}$ is expressed as

$$\chi_\mathrm{R}^{(2)} = \sum_a \frac{A_a}{\omega_a - \omega_\mathrm{MIR} - i\Gamma_a} = \sum_a \left[ \frac{A_a(\omega_a - \omega_\mathrm{MIR})}{(\omega_a - \omega_\mathrm{MIR})^2 + \Gamma_a^2} + i \frac{A_a \Gamma_a}{(\omega_a - \omega_\mathrm{MIR})^2 + \Gamma_a^2} \right], \quad \text{(S49)}$$

where $\omega_\mathrm{MIR}$ is the frequency of incident mid-IR pulses; $\omega_a$ and $\Gamma_a$ represent the resonant frequency and damping constant of vibrational mode $a$, respectively. The first and second terms inside the square brackets correspond to the real and imaginary parts of the second-order nonlinear optical susceptibility derived from the vibrational mode $a$ $\left(\chi_{R,a}^{(2)}\right)$, respectively. Importantly, since $[(\omega_a - \omega_{MIR})^2 + \Gamma_a^2]$ term in the denominator of $\text{Im}\left(\chi_{R,a}^{(2)}\right)$ and $\Gamma_a$ term in the numerator of $\text{Im}\left(\chi_{R,a}^{(2)}\right)$ are always positive, the overall positive/negative sign of $\text{Im}\left(\chi_{R,a}^{(2)}\right)$ is determined by $A_a$. Thus, we next overview the explicit expression of $A_a$ and its relationship with the absolute up/down orientation of 4-MBT molecules.

### 10.2 Relationship between $A_q$ and molecular orientation

Comparing eqs. (S13), (S19), and (S49), $A_q$ is given as[33–35]

$$A_a \propto \sum_{i,j,k}^{x,y,z} \frac{\partial \alpha_{ij}}{\partial u_a} \frac{\partial \mu_k}{\partial u_a} \langle (\hat{Z} \cdot \hat{\imath})(\hat{Z} \cdot \hat{\jmath})(\hat{Z} \cdot \hat{k}) \rangle. \tag{S50}$$

The three directional cosine terms are expressed as $(\hat{Z} \cdot \hat{x}) = -\sin\theta_\ell \cos\psi_\ell$, $(\hat{Z} \cdot \hat{y}) = \sin\theta_\ell \sin\psi_\ell$, and $(\hat{Z} \cdot \hat{z}) = \cos\theta_\ell$, where $\theta_\ell$ and $\psi_\ell$ are the tilt angle and rotational angle of $\ell$-th 4-MBT molecule, respectively (Figure S8b). Moreover, since molecular polarizability is represented by a second-order symmetric tensor, the relation $\partial\alpha_{ij}/\partial u_a = \partial\alpha_{ji}/\partial u_a$ can be applied. Using these relations, we can rewrite the eq. (S50) as follows:

$$A_a \propto \frac{1}{2}\sum_{\ell=1}^{N}\left[(\cos\theta_\ell - \cos^3\theta_\ell)(\beta_{xxz} + \beta_{yyz} + 2\beta_{xzx} + 2\beta_{yzy}) + 2\cos^3\theta_\ell\,\beta_{zzz}\right], \tag{S51}$$

where $\beta_{ijk}$ represents the product of a polarizability derivative and a dipole derivative $\left(\beta_{ijk} = (\partial\alpha_{ij}/\partial u_a)(\partial\mu_k/\partial u_a)\right)$. In deducing eq. (S51) from eq. (S50), we assumed a uniform distribution for $\psi_\ell$ because the activation barrier of methyl rotation is approximately 5 cm$^{-1}$,[22] which is significantly lower than the thermal energy of room temperature (~200 cm$^{-1}$). Thus, the methyl group of 4-MBT can be assumed to be freely rotating under our room-temperature experimental conditions. In this case, the relations of $\sum_{\ell=1}^{N}\cos\psi_\ell = \sum_{\ell=1}^{N}\sin\psi_\ell = \sum_{\ell=1}^{N}\cos^3\psi_\ell = \sum_{\ell=1}^{N}\sin^3\psi_\ell = 0$ and $\sum_{\ell=1}^{N}\cos^2\psi_\ell = \sum_{\ell=1}^{N}\sin^2\psi_\ell = 1/2$ can be applied, making $A_q$ independent of $\psi_\ell$. Moreover, we assume a delta-function distribution for $\theta_\ell$, allowing us to remove the molecular index $\ell$: $\theta_\ell = \theta$. In this situation, $A_a$ is simply described by the tilt angle $\theta$ and the tensor elements $\beta_{ijk}$:

$$A_a \approx \frac{N}{2}\left[(\cos\theta - \cos^3\theta)(\beta_{xxz} + \beta_{yyz} + 2\beta_{xzx} + 2\beta_{yzy}) + 2\beta_{zzz}\cos^3\theta\right]. \tag{S52}$$

This equation describes the relationship between $A_a$ and the molecular tilt angle $\theta$. Notably, $\cos\theta$ and $\cos^3\theta$ in eq. (S52) reflect information of whether hydrogen atoms in the methyl group of 4-MBT is oriented upward or downward relative to the surface (although a downward orientation is not physically realized in the actual SAM). Specifically, for a H-up orientation ($0° < \theta < 90°$), both $\cos\theta$ and $\cos^3\theta$ take positive values, whereas for a H-down orientation ($90° < \theta < 180°$), they take negative values. Since the value of $A_a$ is determined as a linear combination of these terms, the up/down orientation of 4-MBT molecules is encoded in the sign of $A_a$. To elucidate the relationship between the sign of $A_a$ and the absolute orientation, obtaining explicit values of $\beta_{ijk}$ is required. Thus, we next compute the values of $\beta_{ijk}$ through quantum chemical calculations.

### 10.3 Computation of $\beta_{ijk}$

The $\beta_{ijk}$ values for each mode were obtained by calculating both $\partial\alpha_{ij}/\partial u_a$ and $\partial\mu_k/\partial u_a$ values through density functional theory calculations using the B3LYP hybrid functional[13,14] in Gaussian 16 package.[15] Structural optimization and calculation of the tensor elements were performed for a 4-MBT-Au$_3$ cluster (Figure S4b). In the calculation, aug-cc-pVTZ basis sets[16] were used for C, H, and S atoms while Au atoms were described with LanL2DZ basis set and effective core potentials.[17] The tensor elements $\partial\alpha_{ij}/\partial u_a$ and $\partial\mu_k/\partial u_a$ calculated for a 4-MBT-Au$_3$ cluster are displayed in Table S4. Each tensor component of $\beta_{ijk}$ was obtained by taking the product of $\partial\alpha_{ij}/\partial u_a$ and $\partial\mu_k/\partial u_a$, and displayed in Table S5. Note that while all of 27 tensor components of $\beta_{ijk}$ are shown, only five of them involved in eq. (S52) (yellow background in Table S5) are required to obtain the values of $A_a$.

### 10.4 Results

The relationship between $A_a$ and $\theta$ for each vibrational mode, which was calculated by substituting the $\beta_{ijk}$ values in Table S5 into eq. (S52), is presented in Figure S11. For all vibrational modes, $A_a$ is predominantly negative when $0° < \theta < 90°$ (H-up) and positive when $90° < \theta < 180°$ (H-down), indicating that the sign of $A_a$—and hence that of $\text{Im}\left(\chi_{R,a}^{(2)}\right)$—inherently reflects information about the absolute molecular up/down orientation. In our experimental $\text{Im}\left(\chi_{R,a}^{(2)}\right)$ spectra (Figures 3c and d), negative values were indeed obtained for all observed vibrational modes, which is well consistent with H-up configuration of 4-MBT molecules in the SAM at the Au surface.

The difference in the $\theta$ dependence of $A_a$ for each mode can be understood as follows. In the case of symmetric stretching, the absolute value of $\beta_{zzz}$ is significantly larger than other tensor elements (Table S7), making the overall curve predominantly governed by $\cos^3\theta$ in the second term inside the square bracket in eq. (S52) (Figure S11a). In contrast, asymmetric stretching modes exhibit less dominant $\beta_{zzz}$ compared to other tensor elements (Table S7). Consequently, the $\theta$ dependence of $A_a$ largely reflects the $(\cos\theta - \cos^3\theta)$ term in eq. (S52), resulting in a local minimum and maximum at around $\theta = 55°$ and $125°$, respectively (Figure S11b). Finally, in the case of C–H stretching modes in the benzene ring (Figure S11c), the $\theta$ dependence exhibits the intermediate behavior of the symmetric and asymmetric cases owing to the comparable magnitudes of $\beta_{zzz}$ and other tensor elements (Table S7).

Due to the variation in the $\theta$ dependence across the three vibrational modes, the relative ratios of $A_a$ for different modes also exhibit a strong $\theta$ dependence (Figure 6b in the main text and Figure S12). Notably, in the range of $0° < \theta < 90°$, which corresponds to the H-up orientation of the 4-MBT SAM used in this study, all the three intensity ratios ($A_{r^-}/A_{r^+}$, $A_{ph}/A_{r^+}$, and $A_{r^-}/A_{ph}$) increase monotonically. Thus, these curves serve as calibration curves for the estimation of molecular tilt angles $\theta$ from experimentally obtained SFG intensities for each vibrational mode. See Figure 6b in the main text, Figure S12, and Table S8 in the next section for the estimated molecular tilt angles $\theta$.

Table S4. Derivatives of polarizability ($\boldsymbol{\alpha}$) and dipole ($\boldsymbol{\mu}$) tensors for each vibrational mode of a 4-MBT-Au$_3$ cluster. The (3×3) matrix is $\partial\boldsymbol{\alpha}/\partial u_a = \begin{pmatrix} \partial\alpha_{xx}/\partial u_a & \partial\alpha_{xy}/\partial u_a & \partial\alpha_{xz}/\partial u_a \\ \partial\alpha_{yx}/\partial u_a & \partial\alpha_{yy}/\partial u_a & \partial\alpha_{yz}/\partial u_a \\ \partial\alpha_{zx}/\partial u_a & \partial\alpha_{zy}/\partial u_a & \partial\alpha_{zz}/\partial u_a \end{pmatrix}$, and the (3×1) matrix is $\partial\boldsymbol{\mu}/\partial u_a = \begin{pmatrix} \partial\mu_x/\partial u_a \\ \partial\mu_y/\partial u_a \\ \partial\mu_z/\partial u_a \end{pmatrix}$, where $x$, $y$, and $z$ are the molecular fixed coordinates defined in Figure S4b. Unit: atomic units.

| Mode | Polarizability derivatives / $10^{-1}$ a.u. | Dipole derivatives / $10^{-3}$ a.u. |
|---|---|---|
| $r^+$ | $\begin{pmatrix} 1.05 & -0.176 & 0.598 \\ -0.176 & 1.73 & -1.05 \\ 0.598 & -1.05 & 5.02 \end{pmatrix}$ | $\begin{pmatrix} -0.451 \\ 0.598 \\ -3.34 \end{pmatrix}$ |
| $r_1^-$ | $\begin{pmatrix} 0.829 & 0.577 & 0.102 \\ 0.577 & -0.479 & 1.85 \\ 0.102 & 1.85 & 0.100 \end{pmatrix}$ | $\begin{pmatrix} -0.230 \\ -2.18 \\ -1.54 \end{pmatrix}$ |
| $r_2^-$ | $\begin{pmatrix} -0.379 & 0.451 & 1.39 \\ 0.451 & 0.323 & -0.148 \\ 1.39 & -0.148 & 0.409 \end{pmatrix}$ | $\begin{pmatrix} -2.43 \\ -0.006 \\ 0.825 \end{pmatrix}$ |
| $r_{ph1}$ | $\begin{pmatrix} 1.41 & 0.00 & 1.58 \\ 0.00 & 0.065 & -0.032 \\ 1.58 & -0.032 & 1.08 \end{pmatrix}$ | $\begin{pmatrix} -1.77 \\ 0.056 \\ -1.84 \end{pmatrix}$ |
| $r_{ph2}$ | $\begin{pmatrix} 0.943 & 0.054 & -1.22 \\ 0.054 & 0.062 & -0.069 \\ -1.22 & -0.069 & 0.818 \end{pmatrix}$ | $\begin{pmatrix} 1.26 \\ 0.149 \\ -1.75 \end{pmatrix}$ |

Table S5. Hyperpolarizability tensor elements $\left(\beta_{ijk} = (\partial\alpha_{ij}/\partial u_a)(\partial\mu_k/\partial u_a)\right)$ for each vibrational mode of a 4-MBT-Au$_3$ cluster. The elements necessary for the calculation of $A_a$ (eq. (S52)) are highlighted by yellow backgrounds. Unit: $10^{-4}$ atomic units.

| Mode | $r^+$ | $r_1^-$ | $r_2^-$ | $r_{ph1}$ | $r_{ph2}$ |
|---|---|---|---|---|---|
| $\beta_{xxx}$ | −0.473 | −0.191 | 0.922 | −2.50 | 1.19 |
| $\beta_{xxy}$ | 0.626 | −1.80 | 0.002 | 0.079 | 0.140 |
| $\beta_{xxz}$ | **−3.50** | **−1.28** | **−0.313** | **−2.61** | **−1.65** |
| $\beta_{xyx}$ | 0.079 | −0.133 | −1.10 | 0.001 | 0.069 |
| $\beta_{xyy}$ | −0.105 | −1.25 | −0.003 | 0.000 | 0.008 |
| $\beta_{xyz}$ | 0.587 | −0.890 | 0.372 | 0.002 | −0.095 |
| $\beta_{xzx}$ | −0.270 | −0.024 | −3.39 | −2.79 | −1.54 |
| $\beta_{xzy}$ | 0.357 | −0.222 | −0.009 | 0.088 | −0.181 |
| $\beta_{xzz}$ | −2.00 | −0.158 | 1.15 | −2.92 | 2.13 |
| $\beta_{yxx}$ | 0.079 | −0.133 | −1.10 | 0.001 | 0.069 |
| $\beta_{yxy}$ | −0.105 | −1.25 | −0.003 | 0.000 | 0.008 |
| $\beta_{yxz}$ | 0.587 | −0.890 | 0.372 | 0.002 | −0.095 |
| $\beta_{yyx}$ | −0.781 | 0.110 | −0.787 | −0.116 | 0.079 |
| $\beta_{yyy}$ | 1.03 | 1.04 | −0.002 | 0.004 | 0.009 |
| $\beta_{yyz}$ | −5.78 | 0.740 | 0.267 | −0.121 | −0.109 |
| $\beta_{yzx}$ | 0.473 | −0.426 | 0.360 | 0.057 | −0.086 |
| $\beta_{yzy}$ | −0.626 | −4.03 | 0.001 | −0.002 | −0.010 |
| $\beta_{yzz}$ | 3.49 | −2.86 | −0.122 | 0.059 | 0.120 |
| $\beta_{zxx}$ | −0.270 | −0.024 | −3.39 | −2.79 | −1.54 |
| $\beta_{zxy}$ | 0.357 | −0.222 | −0.009 | 0.088 | −0.181 |
| $\beta_{zxz}$ | −2.00 | −0.158 | 1.15 | −2.92 | 2.13 |
| $\beta_{zyx}$ | 0.473 | −0.426 | 0.360 | 0.057 | −0.086 |
| $\beta_{zyy}$ | −0.626 | −4.03 | 0.001 | −0.002 | −0.010 |
| $\beta_{zyz}$ | 3.49 | −2.86 | −0.122 | 0.059 | 0.120 |
| $\beta_{zzx}$ | −2.26 | −0.023 | −0.996 | −1.90 | 1.03 |
| $\beta_{zzy}$ | 3.00 | −0.218 | −0.003 | 0.060 | 0.122 |
| $\beta_{zzz}$ | −16.74 | −0.154 | 0.338 | −1.99 | −1.43 |

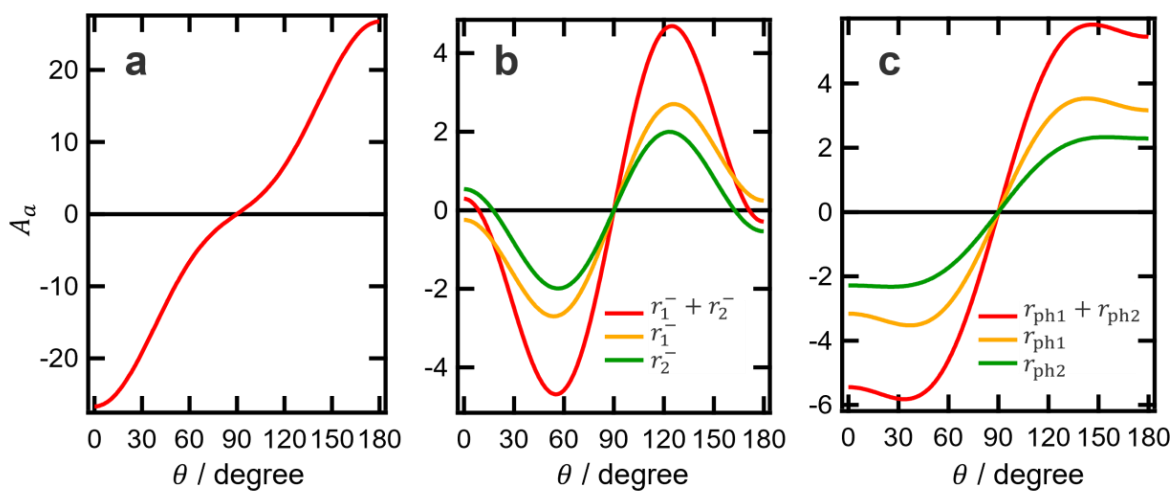

Figure S11. The relationships between $A_a$ and $\theta$ calculated for a 4-MBT-Au$_3$ cluster. The results for methyl symmetric stretching mode (a), methyl asymmetric stretching modes (b), and C–H stretching modes in the benzene ring (c) are displayed. In b, and c, the curves for two quasi-degenerate modes (green and orange curves) and their sum (red curve) are presented.

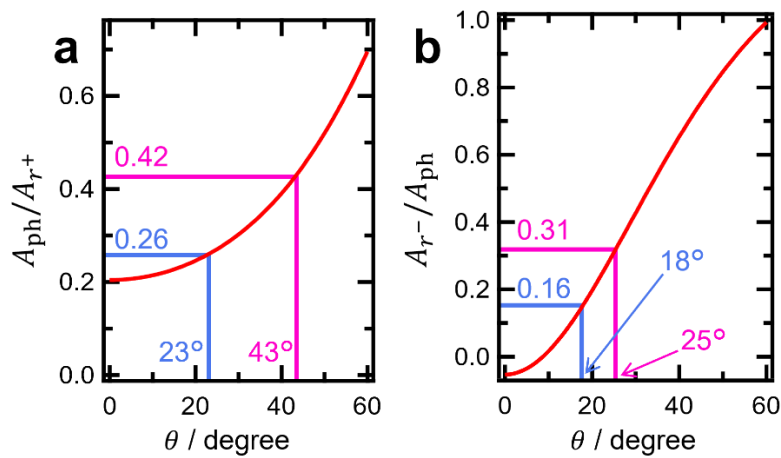

Figure S12. SFG intensity ratios of different vibrational modes calculated for 4-MBT-Au$_3$ cluster. Data are plotted against molecular tilt angles from surface normal ($\theta$). Panels (a) and (b) represent the $\theta$-dependence of $A_{\mathrm{ph}}/A_{r^+}$ and $A_{r^-}/A_{\mathrm{ph}}$, respectively. Blue and magenta horizontal lines indicate the experimentally obtained intensity ratios for rough and flat domains, respectively, and vertical lines indicate the corresponding tilt angles. The estimated values of intensity ratios and tilt angles are indicated in the figures.

## 11. Validity of the calculation: influence of gold attachment on the estimation of molecular tilt angle $\theta$

The quantum chemical calculation presented in the previous section was based on a 4-MBT-Au$_3$ cluster, where three Au atoms are attached to a sulfur atom of a 4-MBT molecule. In this section, we explicitly show the influence of the Au atom attachment to a 4-MBT molecule on the estimation of molecular tilt angle.

We performed an additional quantum chemical calculation for a 4-MBT molecule without Au atoms attachment (Figure S8a), and calculated $\partial \alpha_{ij}/\partial u_a$, $\partial \mu_k/\partial u_a$, and their products ($\beta_{ijk}$) in the same manner as presented in the previous section. The calculated values are shown in Table S6 and Table S7. By substituting $\beta_{ijk}$ values in Table S7 into eq. (S52), we obtained $A_a - \theta$ curves (Figure S13) and three kinds of SFG intensity ratios (Figure S14) for a 4-MBT molecule, from which $\theta$ values for the rough and flat domains were estimated. The values of $\theta$ estimated for a 4-MBT molecule and a 4-MBT-Au$_3$ cluster, as well as their averages and standard deviations, are summarized in Table S8. As shown in Table S8, the three $\theta$ values estimated for a 4-MBT molecule showed poor consistence with large standard deviations comparable to the average $\theta$ values, whereas those for a 4-MBT-Au$_3$ cluster exhibited significantly reduced standard deviations for both rough and flat domains. This indicates that the quantum chemical calculation for a 4-MBT-Au$_3$ cluster provides more physically realistic results than that for a 4-MBT molecule, thereby reinforcing the validity of our analytical approaches shown in the previous section.

Table S6. Derivatives of polarizability ($\alpha$) and dipole ($\mu$) tensors for each vibrational mode of a 4-MBT molecule without Au attachment. The (3×3) matrix is $\partial\alpha/\partial u_a = \begin{pmatrix} \partial\alpha_{xx}/\partial u_a & \partial\alpha_{xy}/\partial u_a & \partial\alpha_{xz}/\partial u_a \\ \partial\alpha_{yx}/\partial u_a & \partial\alpha_{yy}/\partial u_a & \partial\alpha_{yz}/\partial u_a \\ \partial\alpha_{zx}/\partial u_a & \partial\alpha_{zy}/\partial u_a & \partial\alpha_{zz}/\partial u_a \end{pmatrix}$, and the (3×1) matrix is $\partial\mu/\partial u_a = \begin{pmatrix} \partial\mu_x/\partial u_a \\ \partial\mu_y/\partial u_a \\ \partial\mu_z/\partial u_a \end{pmatrix}$, where $x$, $y$, and $z$ are the molecular fixed coordinates defined in Figure S8a. Unit: atomic units.

| Mode | Polarizability derivatives / $10^{-1}$ a.u. | Dipole derivatives / $10^{-3}$ a.u. |
|---|---|---|
| $r^+$ | $\begin{pmatrix} 0.987 & -0.060 & 0.163 \\ -0.060 & 1.70 & -0.644 \\ 0.163 & -0.644 & 3.54 \end{pmatrix}$ | $\begin{pmatrix} -0.185 \\ 0.879 \\ -4.38 \end{pmatrix}$ |
| $r_1^-$ | $\begin{pmatrix} 1.01 & 0.344 & 0.052 \\ 0.344 & -0.249 & 1.54 \\ 0.052 & 1.54 & 0.280 \end{pmatrix}$ | $\begin{pmatrix} -0.069 \\ -2.82 \\ -1.34 \end{pmatrix}$ |
| $r_2^-$ | $\begin{pmatrix} -0.379 & 0.674 & 1.23 \\ 0.674 & 0.232 & -0.001 \\ 1.23 & -0.001 & 0.112 \end{pmatrix}$ | $\begin{pmatrix} -2.90 \\ -0.031 \\ 0.321 \end{pmatrix}$ |
| $r_{ph1}$ | $\begin{pmatrix} 0.887 & 0.039 & 0.978 \\ 0.039 & 0.068 & 0.038 \\ 0.978 & 0.038 & 0.608 \end{pmatrix}$ | $\begin{pmatrix} -1.02 \\ -0.081 \\ -2.48 \end{pmatrix}$ |
| $r_{ph2}$ | $\begin{pmatrix} -0.566 & 0.045 & 1.56 \\ 0.045 & -0.031 & -0.003 \\ 1.56 & -0.003 & -0.380 \end{pmatrix}$ | $\begin{pmatrix} -1.54 \\ 0.033 \\ 1.69 \end{pmatrix}$ |

Table S7. Hyperpolarizability tensor elements $\left(\beta_{ijk} = (\partial\alpha_{ij}/\partial u_a)(\partial\mu_k/\partial u_a)\right)$ for each vibrational mode of a 4-MBT molecule without Au attachment. The elements necessary for the calculation of $A_a$ (eq. (S52)) are highlighted by yellow backgrounds. Unit: $10^{-4}$ atomic units.

| Mode | $r^+$ | $r_1^-$ | $r_2^-$ | $r_{ph1}$ | $r_{ph2}$ |
|---|---|---|---|---|---|
| $\beta_{xxx}$ | −0.183 | −0.070 | 1.10 | −0.902 | 0.875 |
| $\beta_{xxy}$ | 0.868 | −2.84 | −0.012 | −0.072 | −0.019 |
| $\beta_{xxz}$ | −4.33 | −1.35 | −0.122 | −2.20 | −0.956 |
| $\beta_{xyx}$ | 0.011 | −0.024 | −1.96 | −0.039 | −0.069 |
| $\beta_{xyy}$ | −0.053 | −0.968 | 0.021 | −0.003 | 0.001 |
| $\beta_{xyz}$ | 0.265 | −0.459 | 0.217 | −0.096 | 0.075 |
| $\beta_{xzx}$ | −0.030 | −0.004 | −3.56 | −0.994 | −2.42 |
| $\beta_{xzy}$ | 0.143 | −0.146 | 0.038 | −0.080 | 0.051 |
| $\beta_{xzz}$ | −0.713 | −0.069 | 0.394 | −2.42 | 2.64 |
| $\beta_{yxx}$ | 0.011 | −0.024 | −1.96 | −0.039 | −0.069 |
| $\beta_{yxy}$ | −0.053 | −0.968 | 0.021 | −0.003 | 0.001 |
| $\beta_{yxz}$ | 0.265 | −0.459 | 0.217 | −0.096 | 0.075 |
| $\beta_{yyx}$ | −0.314 | 0.017 | −0.674 | −0.070 | 0.048 |
| $\beta_{yyy}$ | 1.49 | 0.701 | 0.007 | −0.006 | −0.001 |
| $\beta_{yyz}$ | −7.44 | 0.332 | 0.075 | −0.170 | −0.052 |
| $\beta_{yzx}$ | 0.119 | −0.107 | 0.003 | −0.039 | 0.004 |
| $\beta_{yzy}$ | −0.567 | −4.34 | 0.000 | −0.003 | 0.000 |
| $\beta_{yzz}$ | 2.82 | −2.06 | 0.000 | −0.094 | −0.005 |
| $\beta_{zxx}$ | −0.030 | −0.004 | −3.56 | −0.994 | −2.42 |
| $\beta_{zxy}$ | 0.143 | −0.146 | 0.038 | −0.080 | 0.051 |
| $\beta_{zxz}$ | −0.713 | −0.069 | 0.394 | −2.42 | 2.64 |
| $\beta_{zyx}$ | 0.119 | −0.107 | 0.003 | −0.039 | 0.004 |
| $\beta_{zyy}$ | −0.567 | −4.34 | 0.000 | −0.003 | 0.000 |
| $\beta_{zyz}$ | 2.82 | −2.06 | 0.000 | −0.094 | −0.005 |
| $\beta_{zzx}$ | −0.655 | −0.019 | −0.325 | −0.619 | 0.587 |
| $\beta_{zzy}$ | 3.11 | −0.789 | 0.003 | −0.050 | −0.013 |
| $\beta_{zzz}$ | −15.50 | −0.374 | 0.036 | −1.51 | −0.642 |

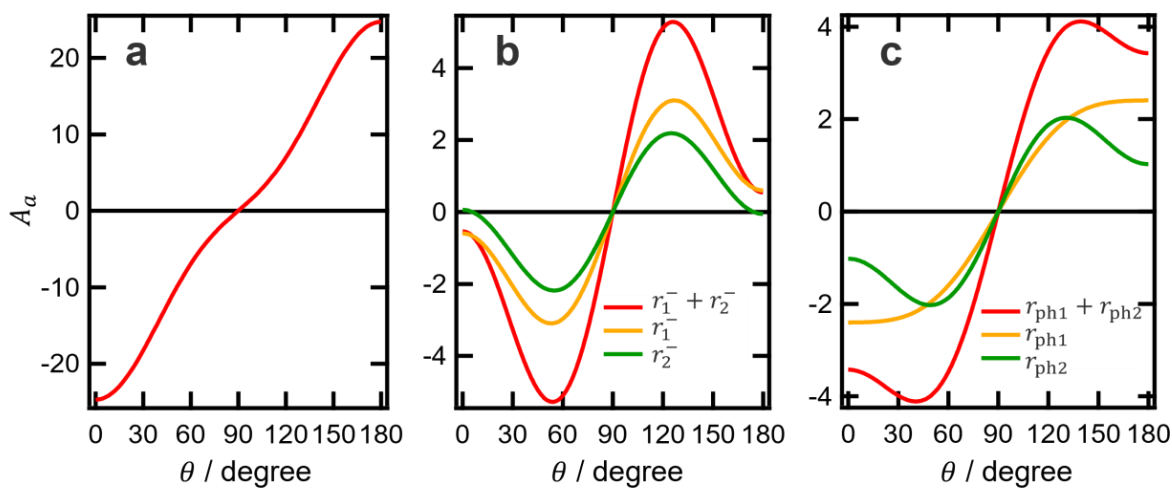

Figure S13. The relationships between $A_a$ and $\theta$ calculated for a 4-MBT molecule without Au attachment. The results for methyl symmetric stretching mode (a), methyl asymmetric stretching modes (b), and C–H stretching modes in the benzene ring (c) are displayed. In b, and c, the curves for two quasi-degenerate modes (green and orange curves) and their sum (red curve) are presented.

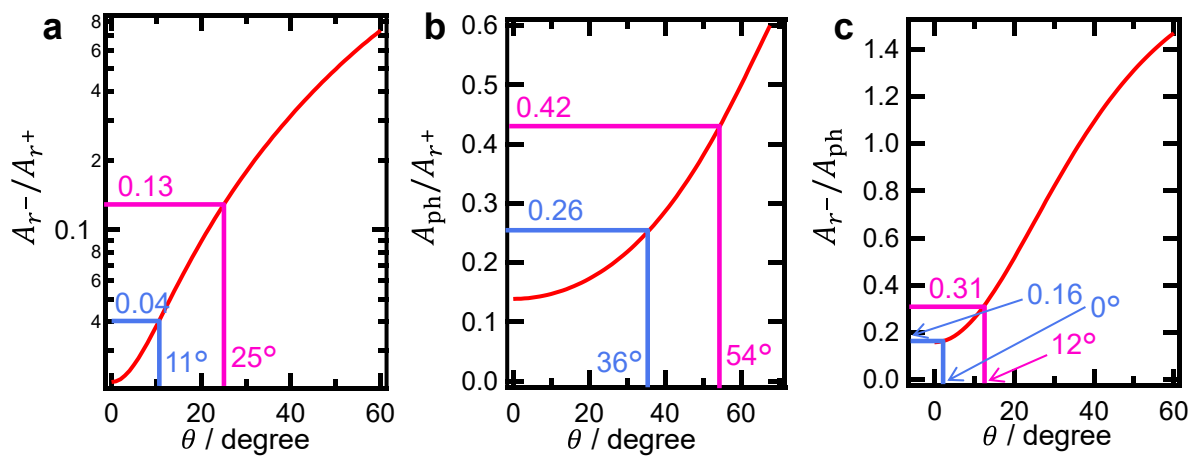

Figure S14. SFG intensity ratios of different vibrational modes calculated for a 4-MBT molecule without an attachment of Au atoms. Data are plotted against methyl tilting angles from surface normal ($\theta$). The left (a), middle (b), and right (c) panels represent the $\theta$-dependence of $A_{r-}/A_{r+}$, $A_{ph}/A_{r+}$, and $A_{r-}/A_{ph}$, respectively. Blue and magenta horizontal lines indicate the experimentally obtained intensity ratios for rough and flat domains, respectively, and vertical lines indicate the corresponding tilt angles. The estimated values of intensity ratios and tilt angles are indicated in the figures.

Table S8. Values of molecular tilt angle $\theta$ estimated from three different calibration curves for a 4-MBT molecule without Au attachment and a 4-MBT-Au$_3$ cluster. Averaged $\theta$ values and their standard deviations are also shown. The $\theta$ values and their errors described in the main text (($20 \pm 2$)° and ($33 \pm 8$)° for the major and minor domains, respectively) are highlighted by yellow backgrounds.

|  | Major domain | | Minor domain | |
| --- | --- | --- | --- | --- |
|  | 4-MBT | 4-MBT-Au$_3$ | 4-MBT | 4-MBT-Au$_3$ |
| $\theta$ estimated from $A_{r-}/A_{r+}$ | 11° | 19° | 25° | 30° |
| $\theta$ estimated from $A_{ph}/A_{r+}$ | 36° | 23° | 54° | 43° |
| $\theta$ estimated from $A_{r-}/A_{ph}$ | 2° | 18° | 12° | 25° |
| Average | 16° | 20° | 30° | 33° |
| Standard deviation | 14° | 2° | 18° | 8° |